\documentclass[aps,twocolumn,nofootinbib,showpacs,floatfix]{revtex4}
\usepackage{graphics} 
\usepackage{amssymb,amsmath}
\usepackage{amsmath}
\usepackage{graphicx}
\tighten

\newcommand{\al}{\alpha} 
 
\newcommand{\de}{\delta} 
 
\newcommand{\ep}{\epsilon}
\newcommand{\ga}{\gamma}

\newcommand{\om}{\omega}

\newcommand{\be}{\begin{equation}} 
\newcommand{\ee}{\end{equation}} 
\newcommand{\bea}{\begin{eqnarray}} 
\newcommand{\eea}{\end{eqnarray}}  
\newcommand{\bean}{\begin{eqnarray*}} 
\newcommand{\eean}{\end{eqnarray*}}

\def\lsim{\raise 0.4ex\hbox{$<$}\kern -0.8em\lower 0.62ex\hbox{$\sim$}} 
\def\gsim{\raise 0.4ex\hbox{$>$}\kern -0.7em\lower 0.62ex\hbox{$\sim$}} 
\newcommand{\bk}{{\bf k}}  
\newcommand{\bx}{{\bf x}}  
\newcommand{\bv}{{\bf v}}
\newcommand{\br}{{\bf r}}

\newcommand{\bn}{{\bf n}} 
\newcommand{\bu}{{\bf u}}
\newcommand{\bR}{{\bf R}}

\newcommand{\bg}{{\mathbf g}}

\newcommand{\tbu}{{\tilde\bu}}

\newcommand{\bq}{{\mathbf q}}
\newcommand{\bse}{\begin{subequations}}
\newcommand{\ese}{\end{subequations}}

\newcommand{\da}{{\dot a}}
\newcommand{\dda}{{\ddot a}}

\newcommand{\bhk}{{\bf \hat k}}

\def\mP{\mathcal{P}}
\def\mQ{\mathcal{Q}}

\begin{document} 
\title{Quantification of discreteness effects in cosmological $N$-body simulations: \\  II. Evolution up to shell crossing
}
\author{M. Joyce}   
\affiliation{Laboratoire de Physique Nucl\'eaire et de Hautes Energies,  
UMR-7585, Universit\'e Pierre et Marie Curie - Paris 6,
75252 Paris Cedex 05, France.} 
\author{B. Marcos}   
\affiliation{ ``E. Fermi'' Center, Via Panisperna 89 A, Compendio del 
Viminale, I-00184 Rome, Italy\\
\& ISC-CNR, Via dei Taurini 29, I-00184 Rome, Italy}

\begin{abstract}   
\begin{center}    
{\large\bf Abstract}   
\end{center}    
  
We apply a recently developed perturbative formalism which describes
the evolution under their self-gravity of particles displaced
from a perfect lattice to quantify precisely, up to shell crossing, the 
effects of discreteness in dissipationless cosmological $N$-body 
simulations. We give simple expressions, explicitly dependent
on the particle density, for the evolution of power in each mode 
as a function of red-shift.  For typical starting 
red-shifts the effect of finite particle number is 
to {\it slow down} slightly the growth of power compared to that in 
the fluid limit (e.g. by about ten percent at half
the Nyquist frequency), and to induce also dispersion in the growth
as a function of direction at a comparable level. 
In the limit that the initial red-shift tends to
infinity, at fixed particle density, the evolution in fact 
diverges from that in the fluid limit (described by the Zeldovich 
approximation). Contrary to widely held belief,  
this means that a simulation started at a red-shift much higher 
than the red-shift of shell crossing actually gives a 
worse, rather than a better, result.
We also study how these effects are modified when
there is a small-scale regularization of the gravitational force.
We show that such a smoothing may reduce the anisotropy of the 
discreteness effects, but it then {\it increases} their average effect.
This behaviour illustrates the fact that the discreteness
effects described here are distinct from those usually
considered in this context, due to two-body collisions.
Indeed the characteristic time for divergence from
the collisionless limit is proportional to
$N^{2/3}$, rather than $N/ \log N$ in the latter case.

\end{abstract}    
\maketitle   
\date{today}  

\twocolumngrid   

\section{Introduction}

Cosmological $N$-body simulations have become an essential
instrument in the attempt to understand the origin of large
scale structure in the universe within the framework of
current cosmological models (see, e.g., 
\cite{efstathiou_init, couchman, springel_05}). In this context 
the goal is to use such simulations to recover the clustering of dark matter,
which is described by a set of coupled Vlasov-Boltzmann equations
corresponding to an appropriately taken $N \rightarrow \infty$ 
limit. The problem of discreteness is that of the relation
between these finite $N$ simulations and this continuum limit.
It is a problem which is becoming rapidly of greater importance
because of the need for great precision in the predictions furnished
by simulations, as the observations constraining them will continue
to improve in the coming years (see e.g. \cite{Huterer:2004tr}
for a discussion of precision requirements for future weak lensing 
surveys). 

This article is the second in a series in which we 
address this issue in a quantitative manner. In the first 
paper \cite{discreteness1_mjbm} we have considered
only the initial conditions (IC) for such simulations, studying in detail
the correlation properties of the point distributions generated
with the standard algorithm used to produce them. We have identified
the limit in which the continuum IC are recovered, and quantified
precisely the corrections to this limit which appear at finite 
particle number. We have noted that there is a subtlety about 
how this limit is taken: the exact theoretical correlation 
properties of the IC in real and reciprocal space are recovered only
when the limit $N \rightarrow \infty$ is taken {\it before} the 
limit that the initial redshift $z_{\rm init} \rightarrow \infty$. If,
on the other hand, $z_{\rm init}$ is increased arbitrarily at fixed particle 
density, the real space correlation properties (e.g. mass variance
in spheres) become dominated by those of the underlying 
(unperturbed) particle distribution, and thus differ completely 
from those of the theoretical IC. As the analysis of \cite{discreteness1_mjbm}
is limited to the IC, no conclusion can be drawn about the relevance
of these behaviours in the evolved simulations. One of our results
here is that this non-interchangeability of these limits also
manifests itself in the perturbative approximation to the 
evolution we use. As explained briefly in the conclusion of 
\cite{discreteness1_mjbm}, this behavior can be understood as
a manifestation of the fact that the Vlasov limit of an 
$N$-body system with long-range interactions is valid for
sufficiently short times \cite{braun+hepp,spohn}. More
specifically, this means that taking $z_{\rm init} \rightarrow \infty$,
at fixed particle density, the evolution of the particle system
diverges from that of the Vlasov-Poisson limit. Practically
this implies that increasing the starting red-shift $z_{\rm init}$
of a simulation, keeping all else fixed, the results get worse, in 
the sense that they deviate further from the desired physical 
behaviour. This finding corrects a widespread belief (see, e.g., 
\cite{scoccimarro_transients_98, valageas_IC_03, power_03, TatekawaIC_07})
that the opposite is true, as the only envisaged error has been 
that due to non-linear corrections in the fluid limit (which vanish
when $z_{\rm init} \rightarrow \infty$). 

In this paper we pursue the study of \cite{discreteness1_mjbm} of 
discreteness effects, considering the early times dynamics of these 
simulations. By ``early times''
we mean the regime of validity of a perturbative treatment 
of the evolution of the full $N$-body problem which has been
reported in detail in \cite{joyce_05, marcos_06}. This corresponds
to a time when the relative displacement of particles at adjacent
lattice sites (which is always small compared to unity in the 
IC) become of order the lattice spacing, i.e., when pairs of 
particles first approach one another. In fluid language it
corresponds approximately to ``shell crossing''.  The basis
of the approach is a simple perturbative expansion, in the
relative displacements of pairs of particles, of the force
on each particle. This is in fact a standard method used to
analyze phonons in a crystal, well known in solid state
physics (see e.g. \cite{pines}). It leads to a
3N$\times$3N diagonalisation problem,
which can be simplified using the translational symmetry of
the lattice (formulated in the Bloch theorem): the eigenmodes 
of the displacement
field are simply plane waves, with orientations and eigenvalues
which can be determined numerically by diagonalising 
N 3$\times$3 matrices. Analysing this spectrum in the 
continuum limit of infinite particle density, one recovers 
the expected fluid behaviour in this regime. The latter is 
that derived through a perturbative treatment of the fluid
equations in the Lagrangian formalism \cite{buchert2}, 
which gives asymptotically the Zeldovich approximation.
We will refer to the discrete perturbative treatment here as 
``particle  linear theory'' (PLT) and to its fluid limit as
``fluid linear theory'' (FLT). By comparing evolution described
by the former with the latter, we determine the discreteness
effects (in the domain of validity of PLT).

That this PLT formalism can be used to this end has already been
clearly demonstrated in \cite{marcos_06}. Indeed in this paper
we have directly compared $N$-body simulations with both PLT and
FLT and shown that the former reproduces better than the latter
the real evolution. Further we have shown that PLT reproduces
the full evolution extremely well at early times, and breaks
down globally at roughly the same time as FLT, when the 
average relative displacement of particles becomes of order
the interparticle distance.

In the present paper we apply this formalism in detail to quantifying
discreteness effects in cosmological $N$-body simulations. To our knowledge
this is the first work in which any such effect in simulations has
been quantified using analytic methods. Discreteness effects have, on
the other hand, been investigated, mostly numerically, by various
authors (see e.g. \cite{splinter, kuhlman, melott_alone, melott_all,
diemandetal_convergence, diemandetal_2body, binney_discreteness,
discreteness-hamana,Baertschiger:2002tk, gotz+sommerlarsen_WDM, 
power_03, wang+white_HDM}). While our results are limited to the effects 
which are at play up to shell crossing, they provide a complete and 
exhaustive understanding of this regime. As we discuss in our 
conclusions the physical insights gained should also be useful in attempting
to understand the effects of discreteness better beyond this regime.
Our results also provide an analytical description for some
specific effects of discreteness which have been observed
numerically, notably the effects of discreteness in breaking
isotropy demonstrated by a numerical experiment in
\cite{melott_all}, and the generation of structure at small
scales noted in hot/warm dark matter simulations in 
\cite{gotz+sommerlarsen_WDM, wang+white_HDM}.

The paper is organized as follows. In 
the next section we summarize the necessary formalism which
has been developed and studied in detail in  
\cite{joyce_05,marcos_06}. 
This section is divided into three
parts: the PLT formalism in full generality, the derivation of the 
fluid limit from it, and the specific case of evolution in an
Einstein de Sitter (EdS) universe from the IC applied in 
cosmological simulations (using the Zeldovich approximation). 
For the latter case (i.e. an EdS universe) all of the cosmology 
dependent part of the calculations can be done analytically. It is 
in fact the appropriate case for almost all cosmological applications, 
since we are treating in practice the evolution in an epoch 
(before shell crossing)  which corresponds to a red-shift range 
in which the universe, in current cosmological models, is very close to EdS.
The generalization to any other cosmology (e.g. with a cosmological 
constant) is, however, straightforward. In 
Sect.~\ref{Statistical measures of discreteness effects} we then
apply these results to quantify more precisely the effects in
cosmological simulations. Specifically we calculate a simple
function quantifying the modification of the average amplitude 
of all modes of the displacement field at given wavenumber,
then a similar one for the modes of the density fluctuations
and finally one quantifying the anisotropy induced in the 
evolution by discreteness. In the following section we 
study the case that there is a smooth regularization of the 
Newtonian potential around the origin. In particular we consider
the modification induced in the quantities defined and calculated
in the previous section. In 
Sect.~\ref{Parametric and limiting behaviours} we
consider the parametric and limiting behaviours of the discreteness
effects which we have quantified. In particular we consider more 
explicitly how the effects depend
on the number of particles and the recovery of the continuum limit,
as well as the limit in which the initial amplitude goes to
zero (i.e. the initial red-shift goes to infinity).  
In the final section we summarize our findings
and discuss some other points. In particular we discuss the
possible use of our results to correct cosmological simulations
for these systematic errors due to discreteness at shell crossing.
We also discuss briefly the relevance of our results to the
more general problem of quantifying discreteness errors in the
fully non-linear regime of these simulations. 


\section{PLT Evolution of a perturbed lattice}
\label{PLT Evolution of a perturbed lattice}

In the first part of this section we present the essential elements
of the PLT treatment of the early time evolution of a perturbed
lattice, summarizing a much more detailed discussion which can be
found in \cite{joyce_05, marcos_06}. In the next subsection we
discuss the derivation of the fluid limit of the evolution 
described by PLT.
In the last part we then consider the specific case of evolution
from a lattice as perturbed initially in cosmological simulations.

\subsection{Summary of PLT formalism}
\label{PLT}

The equation of motion of the particles in a dissipationless 
cosmological $N$-body system is (see, e.g., \cite{efstathiou_init, couchman})
\be
{\ddot {\bf x} }_i +
2 H (t) {\dot {\bf x} }_i
= -\frac{1}{a^3} \sum_{i\neq j} 
\frac{G m_j ({\bf x}_{i}-{\bf x}_j) }{|{\bf x}_{i}-{\bf x}_{j}|^3}\,.
\label{eom}
\ee
Here dots denote derivatives with respect to time $t$,
${\bf x}_i$ is the comoving position of the $i$th particle,
of mass $m_i$, related to the physical coordinate by 
${\bf r}_i=a(t) {\bf x}_i$, where $a(t)$ is the scale factor 
of the background cosmology with Hubble constant
$H(t)=\frac{\dot a}{a}$. The infinite universe\footnote{The infinite
sum for the force, as written in Eq.~(\ref{eom}), is formally 
not well defined. It is implicit that it is regularized
by the subtraction of the effect of the mean density, 
the effect of which is taken into account in the expansion 
of the universe. 
In \cite{marcos_06} we have studied the case of a static universe
(i.e, the case where the scale factor
$a(t)=1$, in which the same 
regularization of this sum is applied) more extensively as it 
makes the comparison with the analogous
condensed matter system much more transparent. The differences 
between the static and expanding case are, in any case, 
quite trivial for the PLT approximation.} is treated
by applying periodic boundary conditions to a finite cubic box of
side $L$.

We treat first the generic case that the $N$ particles in 
the box are of equal mass and  initially placed in any 
configuration which is a perturbed lattice, i.e., particles 
initially at the lattice sites of a perfect lattice subjected
to some set of displacements. The right hand side of 
Eq.~(\ref{eom}) may then be expanded perturbatively 
in the relative displacements of particles, about the
perfect lattice configuration in which it is identically
zero. Writing ${\bf x}_i(t)={\bf R} + {\bf u}({\bf R},t)$, 
where ${\bf R}$ is the lattice vector of the $i$th particle
and ${\bf u}({\bf R},t)$ its displacement from {\bf R}, 
one obtains then, at linear order in this expansion, 
\be
{\bf {\ddot u}}({\bf R},t) 
+2 H {\bf {\dot u}}({\bf R},t) 
= -\frac{1}{a^3} \sum_{{\bf R}'} 
{\cal D} ({\bf R}- {\bf R}') {\bf u}({\bf R}',t)\,. 
\label{linearised-eom}
\ee
The matrix ${\cal D}$ is known in solid
state physics, for any pair interaction, as the
{\it dynamical matrix} (see e.g. \cite{pines}).  
For gravity we have 
\bea
{\cal D}_{\mu \nu} ({\bf R} \neq {\bf 0})=
Gm\left(\frac{\delta_{\mu \nu}}{R^3}
-3\frac{R_\mu R_\nu}{R^5}\right) \\
{\cal D}_{\mu \nu} ({\bf 0})= 
-\sum_{{\bf R} \neq {\bf 0}} {\cal D}_{\mu \nu} ({\bf R})
\eea
where $\delta_{\mu \nu}$ is the Kronecker delta.
Note that a sum over the copies, associated with
the periodic boundary conditions, is implicit in
these expressions.
 
The Bloch theorem for lattices tells us that ${\cal D}$
is diagonalized by plane waves in reciprocal space.
We can define the Fourier transform and its inverse by 
\bse
\label{def-discreteFT}
\begin{align}
\label{def-discreteFT-tok}
{\bf {\tilde u}}({\bf k},t)&= \sum_{{\bf R}} e^{-i {\bf k}\cdot{\bf R}}
{\bf u}({\bf R},t) \\
\label{def-discreteFT-tor}
{\bf u}({\bf R},t)&= \frac{1}{N} \sum_{{\bf k}} e^{i {\bf k}\cdot{\bf R}}
 {\bf {\tilde u}}({\bf k},t)\,,
\end{align}
\ese where the sum in Eq.~\eqref{def-discreteFT-tor} is over the first
Brillouin zone (FBZ) of the lattice, i.e., for a simple cubic 
lattice\footnote{The FBZ, for any cubic lattice,
is defined as the set of $N$ non-equivalent reciprocal lattice 
vectors closest to the
origin $\bk=\mathbf 0$.  See \cite{marcos07} for the generalisation of 
the treatment described here to the case of a face centered cubic
and body centered cubic lattice.} $\bk = \bn (2\pi/L)$, where $\bn$
is a vector of integers of which each component $n_i$ ($i=1,2,3$) 
takes all integer values in the range $-N^{1/3}/2 < n_i \leq N^{1/3}/2$. 

Using these definitions in Eq.~(\ref{linearised-eom}) we
obtain \be {\bf \ddot{{\tilde u}}} ({\bf k},t) + 2 H (t) {\bf
\dot{{\tilde u}}} ({\bf k},t) = -\frac{1}{a^3} {\cal {\tilde D}} ({\bf
k}) {{\bf {\tilde u}}}({\bf k},t) \ee where ${\cal {\tilde D}} ({\bf
k})$, the Fourier transform (FT) of ${\cal D} ({\bf R})$, is a
symmetric $3 \times 3$ matrix for each ${\bf k}$.

Diagonalising ${\cal {\tilde D}} ({\bf k})$ one can determine, 
for each ${\bf k}$,  three orthonormal eigenvectors 
${\bf e}_n ({\bf k})$ and their eigenvalues 
$\omega_n^2({\bf k})$ ($n=1,2,3$). The latter obey a sum
rule\footnote{This rule is known in the context of condensed 
matter physics \cite{pines} as the Kohn sum rule.}
\be
\sum_n \omega_n^2({\bf k}) = -4 \pi G \rho_0 \,,
\ee
where $\rho_0$ is the mean mass density. 

Given the initial displacements and velocities 
at a time $t=t_0$, the dynamical evolution of the
particle trajectories is then given as  
\be
\label{eigen_evol}
\bu(\bR,t)=\frac{1}{N}\sum_{\bk}\left[\mathcal{P}(\bk,t)\tbu(\bk,t_0)+\mathcal{Q}(\bk,t)\dot\tbu(\bk,t_0)\right]e^{i\bk\cdot\bR}
\ee
where the matrix elements of the ``evolution operator'' $\mathcal{P}$ and $\mathcal{Q}$ are
\bse
\label{evol_operators}
\begin{align}
\mP_{\mu\nu}(\bk,t)=&\sum_{n=1}^3 U_n(\bk,t)(\mathbf{e}_n(\bk))_\mu(\mathbf{e}_n(\bk))_\nu\\
\mQ_{\mu\nu}(\bk,t)=&\sum_{n=1}^3 V_n(\bk,t)(\mathbf{e}_n(\bk))_\mu(\mathbf{e}_n(\bk))_\nu\,.
\end{align}
\ese
The functions $U_n({\bf k},t)$ and $V_n({\bf k},t)$ are linearly
independent solutions of the mode equations \be {\ddot{f}} + 2 H
{\dot{f}}= -\frac{\omega_n^2({\bf k})}{a^3} f
\label{mode-equation}
\ee
chosen such that
\bea
\nonumber
U_n({\bf k},t_0)=1\,,\,\,\,\dot{U}_n({\bf k},t_0)=0\,,\\ 
V_n({\bf k},t_0)=0\,, \,\,\, \dot{V}_n({\bf k},t_0)=1\,.
\label{normalization}
\eea 
The determination of the evolution thus reduces to:
\begin{itemize}
\item 1. diagonalization of the N 3$\times$3 matrices 
${\cal {\tilde D}} ({\bf k})$, and
\item 2. solution of the 3N equations (\ref{mode-equation})
and (\ref{normalization}) for the mode
functions.
\end{itemize}
We note that ${\cal {\tilde D}} ({\bf k})$, and 
therefore the first step, depends only on the type of
lattice, and {\it not} on the cosmology. The dependence on the latter
[encoded in $a(t)$] comes only in the second step, in the 
mode functions $U_n(t)$ and $V_n(t)$. Both steps are numerically
straightforward (and require a number of operations proportional
to $N$, quite feasible even with moderate computational
power for $N$ as large as those used in the very largest 
current cosmological simulations). For certain cases, notably
an EdS universe, the mode functions can be
written analytically (see \cite{marcos_06}). For completeness
we give these expressions in App.~\ref{appendix-EdSmodefunctions}.

\begin{figure}
\resizebox{8cm}{!}{\includegraphics*{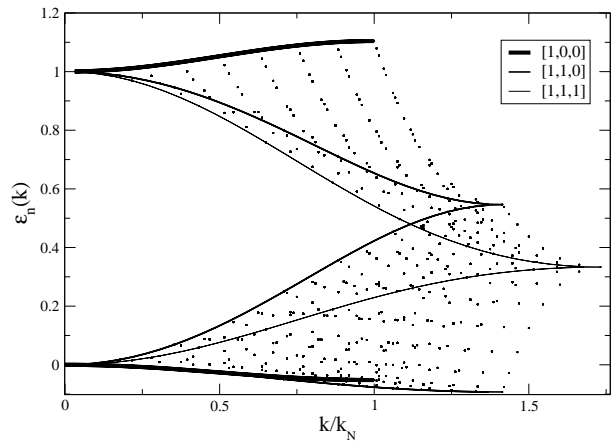}}
\caption{Normalized eigenvalues (see text) for the eigenvectors 
of the  displacement field on a simple cubic lattice with lattice 
spacing $\ell$. Each point corresponds to an eigenvector
labelled by a vector $\bk$ in the first Brillouin zone
of the lattice, and $k_N$ is the  Nyquist frequency. The
different continuous lines link $\bk$ vectors oriented in
the indicated directions. Note that for the $[1,O,O]$ and
$[1,1,1]$ directions the two tranverse eigenmodes are 
degenerate because of rotational symmetry.
\label{fig1}}
\end{figure}

Details of the diagonalization of ${\cal {\tilde D}} ({\bf k})$ are
given in \cite{marcos_06}. In Fig.~\ref{fig1} are shown results for a
$16^3$ simple lattice. Specifically it
shows the normalized eigenvalues
\be \epsilon_n ({\bf k})\equiv -\frac{\omega_n^2({\bf k})}{4\pi G
\rho_0}
\label{def-epsilon}
\ee
where $\rho_0$ is the mean mass density (arbitrarily chosen at the
time $t_0$). These are plotted as a function of the modulus $k$ of
$\bk$, in units of the Nyquist frequency $k_N=\frac{\pi}{\ell}$ where
$\ell$ is the lattice spacing. The fact that the eigenvalues at a
given $k$ do not have the same value is a direct result of the 
anisotropy of the lattice. Shown in
the plot are also lines linking eigenvectors oriented in some chosen
directions. This allows one to see the branch structure of the
spectrum, which is familiar in the context of analogous calculations
in condensed matter physics (see e.g. \cite{pines}). The vectors
$\bk$ are those in the first Brillouin zone of the simple cubic
lattice (as defined above).

The number of particles in the simple cubic lattice considered 
for the above calculation is small compared to that in current
cosmological $N$-body simulations, which now attain (e.g. \cite{springel_05})
values surpassing $1000^3$. While it is quite feasible, as noted above,
to do these calculations for such large particle numbers, it is of
little interest to do so here: the effects we are interested in depend
essentially on the {\it ratio} of the wavenumber of a given mode to the 
Nyquist frequency. Increasing particle number only changes the density
of reciprocal lattice vectors in these units (since $\bk/k_N =
\bn/N^{1/3}$), just filling in more densely the plot of the eigenvalues
in Fig.~\ref{fig1} but leaving its form essentially unchanged.

In \cite{marcos_06} we have investigated the domain of validity of 
PLT using numerical simulations, comparing the results of the evolution
under full gravity with that obtained using the formulae
just given. We have considered  ``shuffled lattice''  IC (in which a 
random uncorrelated displacement is applied
to each particle of the lattice, corresponding to a power spectrum
$P(k) \propto k^2$ at small $k$), and cosmological type IC (a power 
spectrum $\propto k^{-2}$). In both cases 
the full evolution is observed to be very well approximated until 
a time when the average relative displacements approach the 
lattice spacing. 

\subsection{Derivation of the fluid limit (Zeldovich Approximation)}
\label{The Zeldovich Approximation}

The continuum limit of this discrete analysis is easily obtained as
follows. We consider the eigenmodes labelled by ${\bf k}$. 
Sending the particle density to infinity, at fixed mean mass 
density $\rho_0$, we approach the asymptotic behaviour
seen in Fig.~\ref{fig1} as $k/k_N \rightarrow 0$. It is simple 
to show \cite{marcos_06} that one has then one purely longitudinal
mode (i.e. parallel to {\bf k}) with 
$\epsilon_n({\bf k})=1$, and two tranverse modes
with $\epsilon_n({\bf k})=0$. The evolution Eq.~(\ref{eigen_evol}) 
may then be written
\bea
\label{sol_fluid}
\bu(\bR,t) &=& \bu_\perp(\bR,t_0) U_\perp(t) +\bu_\parallel(\bR,t_0)
U_\parallel(t)\\ \nonumber
&&+\bv_\parallel(\bR,t_0) V_\parallel(t)
+ \bv_\perp(\bR,t_0) V_\perp(t),
\eea
where
\bse
\begin{align}
\bu_\parallel(\bR,t_0)&=\frac{1}{N}\sum_{\bk}(\tbu(\bk,t_0)\cdot\hat\bk)\hat\bk\
\,e^{i\bk\cdot\bR},\\
\bu_\perp(\bR,t_0)&=\frac{1}{N}\sum_{\bk} [\tbu(\bk,t_0)-(\tbu(\bk,t_0)
\cdot \hat\bk) \hat\bk ]\,e^{i\bk\cdot\bR},
\end{align}
\ese
and analogously for the velocity fields, i.e., the longitudinal and
transverse components of the initial displacement and velocity fields. 
The functions $U_\parallel(t)$ and $V_\parallel(t)$ are the linearly
independent solutions of Eq.~(\ref{mode-equation}) for 
$\epsilon_n({\bf k})=1$ (i.e. $\omega_n^2({\bf k})=-4\pi G \rho_0$)
), and $U_\perp(t)$ and
$V_\perp(t)$ those for $\epsilon_n({\bf k})=1$ (i.e. $\omega_n^2({\bf k})=0$). We have observed
in \cite{joyce_05} that, for an EdS universe, this corresponds 
exactly to the solution found at linear order in a perturbative
treatment of the equations for a self-gravitating fluid in
the Lagrangian formalism in \cite{buchert2}. 

The solutions for the mode functions give, generically, a 
growing mode and a decaying mode. Since the former always
dominates after a sufficient time, there is an
attractive asymptotic behaviour for the general solution
Eq.~(\ref{sol_fluid}) which may be written in the form
\be
{\bf u}({\bf R}, t) = g(t) {\bf q}({\bf R})\,,
\label{ZA}
\ee
where $g(t)$ is proportional to the purely growing mode,
and ${\bf q}({\bf R})$ is a longitudinal vector field.
This is the Zeldovich approximation (ZA) \cite{zeldovich_70}.

\subsection{Evolution from ZA initial conditions}

In cosmological simulations the ZA is used to fix the IC, usually 
at a time at which the universe is well
approximated by an EdS cosmology. In this case we 
have (see App.~\ref{appendix-EdSmodefunctions}):

\begin{eqnarray}
\label{eds-mode-function}
U_\parallel(t) &=&\frac{2}{5}\left[\frac{3}{2}\left(\frac{t}{t_0}\right)^{2/3}+\left(\frac{t}{t_0}\right)^{-1}\right]\\
V_\parallel(t)&=&\frac{3}{5}t_0\left[\left(\frac{t}{t_0}\right)^{2/3}-\left(\frac{t}{t_0}\right)^{-1}\right]\\
U_\perp(t)&=&1\,,\qquad  V_\perp(t)=3t_0\left[1-\left(\frac{t}{t_0}\right)^{-1/3}\right]\,.
\end{eqnarray}
The ZA may then be written
\bse
\label{zeldo-rel}
\begin{align}
\bu(\bR,t)&=\frac{3}{2}\left(\frac{t}{t_0}\right)^{4/3} \bg(\bR,t_0)t_0^2\\
\bv(\bR,t)&=\bg(\bR,t_0) t,
\end{align}
\ese
where $\bg (\bR,t)$ and $\bv(\bR,t)$ are the  peculiar gravitational
acceleration field and peculiar velocity field, respectively, defined
by 
\bea
\label{peculiar-v+g}
\bv&=&\dot\br-\da\bx\\
\bg&=&\ddot\br-\dda \bx=\ddot\br-\frac{\dda}{a}\br=a\left[\ddot \bu+2\frac{\da}{a}\dot\bu\right]\,.
\eea
We therefore have
\bse
\label{zeldo-cond}
\begin{align}
\bu_\perp(\bR,t_0)&=\mathbf 0=\bv_\perp(\bR,t_0)\\
\bv_\parallel(\bR,t_0)&=\frac{2}{3t_0}\bu_\parallel(\bR,t_0)\,.
\end{align}
\ese
Using these specific IC in the general expression 
for the PLT evolution given in Eq.~(\ref{eigen_evol}), it is simple
to show that this evolution may be written in the form
\be
\label{eigen_evol_cosm}
\bu(\bR,t)=\frac{1}{N}\sum_{\bk} {\bf E} ({\bf k},t)
{\tilde u} ({\bf k},t_0) e^{i\bk\cdot\bR}
\ee
%
where we have defined 
\be
{\bf \tilde u} ({\bf k},t_0) = 
{\bf \hat k} ({\bf \hat k} \cdot {\bf \tilde u} ({\bf k},t_0))
\equiv {\bf \hat k} {\tilde u} ({\bf k}, t_0)\,,
\ee
using the fact that the initial displacement field 
is purely longitudinal. The set of {\it vectors} 
${\bf E} ({\bf k},t)$ which encode the evolution
in PLT of each mode of the displacement field 
are given by 
\be
{\bf E} ({\bf k},t)=\sum_{n}
[U_n({\bf k},t) + \frac{2}{3t_0} V_n({\bf k}, t)]
({\bf \hat k} \cdot {\bf{\hat e}}_n ({\bf k})) {\bf{\hat e}}_n ({\bf k})\,. 
\label{evolution-ZA-E}
\ee
These can be calculated straightforwardly for the given lattice 
and cosmology. As emphasized above, the latter only enters in
determining the mode functions, while the eigenvectors and eigenvalues
are those of the simple cubic lattice widely used in cosmological 
simulations.

The fluid limit is expressed as 
\be
\label{fluid-limit}
{\bf E}^{\rm fluid} ({\bf \hat k},t)= 
\lim_{k \rightarrow 0} {\bf E} (\bk,t)= 
[U_\parallel(t) + \frac{2}{3t_0} V_\parallel (t)]
{\bf \hat k}
\ee
where, as discussed above, the limit is taken at fixed lattice spacing 
$\ell$, and $U_\parallel(t)$ and $V_\parallel (t)$ are the solutions
of Eqs.~(\ref{mode-equation}) and (\ref{normalization}) for
$\epsilon_n(\bk)=1$ (i.e. $\omega_n^2(\bk)=-4\pi G \rho_0$). 
Thus we recover again the ZA, in 
which the initial displacements are simply amplified by the 
appropriate time dependent factor.

From now on we will consider specifically the case only of
the EdS cosmology. Since all the calculations we do with
PLT are valid (and will be applied) only until the time of 
shell-crossing, this means we assume only that the cosmological
model studied is well approximated by EdS from the starting
red-shift until shell-crossing. This is almost always the case,
notably in that of the currently favored $\Lambda CDM$ model,
as the starting red-shift is always well within the matter
dominated era and the cosmological constant contributes to
the expansion significantly only at very low red-shift, 
well after shell crossing of the smallest included
non-linear scales. For this case it is simple to verify,
following the discussion in the previous subsection, that
\be
{\bf E}^{\rm fluid} ({\bf \hat k},t)= \left( \frac{t}{t_0} \right)
^{\frac{2}{3}}\, {\bf \hat k} =a (t)\, {\bf \hat k}
\ee
where $a=1$ at $t=t_0$. 

At sufficiently long times we note that the expression 
Eq.~(\ref{evolution-ZA-E}) will always be dominated 
by the most rapidly growing of the 
three modes at any given $\bk$, 
associated to the largest of three eigenvalues. Using the
expressions in App.~\ref{appendix-EdSmodefunctions} for
the EdS mode functions we can write 
\bea
{\bf E} (\bk,t \gg t_0) \approx&& 
\frac{1}{3}\frac{2+3\al_{\rm max}^+(\bk)}{
(\alpha^-_{\rm max}(\bk)+ \alpha^+_{\rm max}(\bk)) } 
\left( \frac{t}{t_0} \right)^{\alpha^-_{\rm max}(\bk)}\nonumber\\
&\times&  ({\bf \hat k} \cdot {\bf{\hat e}}_{\rm max} ({\bf k})) 
\,{\bf{\hat e}}_{\rm max} ({\bf k})
\label{E-asymptotic}
\eea
where the coefficients $\alpha^\pm_{\rm max}(\bk)$ are those 
given by the expressions in Eqs.~(\ref{alpha-epsilon})
for the largest of the three eigenvalues $\epsilon_n(\bk)$,
with corresponding eigenvector ${\bf{\hat e}}_{\rm max} ({\bf k})$
\footnote{We assume that 
${\bf \hat k} \cdot {\bf{\hat e}}_{\rm max} (\bk) \neq 0$, which is
indeed the case for all $\bk$.}.

\section{Statistical measures of discreteness effects}
\label{Statistical measures of discreteness effects}

As we have explained, discreteness effects may be identified, in
the regime of validity of PLT, as the differences between the 
evolution described by PLT, and the fluid limit (FLT). For
cosmological IC the effect of discreteness on any 
{\it individual} particle trajectory may thus be written as 
\be
\label{disc_cosm}
\Delta \bu_{\rm disc} (\bR,t)=
\frac{1}{N}\sum_{\bk}\left[{\bf E} ({\bf k},t)-{\bf E}^{\rm fluid} (\bhk,t) \right]
{\tilde u} ({\bf k},t_0) e^{i\bk\cdot\bR}
\ee
where the explicit expressions for the 
vectors ${\bf E} (\bk,t)$ and ${\bf E}^{\rm fluid} (\hat\bk,t)$ have
been given above.

To quantify effects of discreteness in simulations we consider 
some statistical measures of the induced effects.

\subsection{The power spectrum of displacements}
\label{Discreteness effects in the power spectrum of displacements}

The two point properties of the displacement field are specified 
by the matrix
\be
{\cal S}_{\mu \nu} ({\bf k}, t)
\equiv {{\tilde u}_\mu}({\bf k},t){{\tilde u}_\nu^*}({\bf k},t).
\ee
When the IC are set up, as just discussed, using the 
ZA, it follows from Eq.~(\ref{eigen_evol_cosm}) that
\be
\label{evolution-Pdisp}
{\cal S}_{\mu \nu} ({\bf k}, t)
= E_\mu({\bf k},t) E_\nu ({\bf k},t) |{\tilde u}({\bf k},t_0)|^2
\ee
and 
\be
{\cal S}_{\mu \nu} ({\bf k}, t_0)
={\hat k}_\mu {\hat k}_\nu |{\tilde u}({\bf k},t_0)|^2
\ee
where we used the fact that ${\bf E}({\bf k},t)$ is real
[and ${\bf E} ({\bf k},t_0)=1$]. 

Let us consider the evolution of the trace of this matrix 
$P_D(\bk,t) \equiv Tr\, {\cal S}({\bf k}, t)$, for which we 
have that  
\be
P_D(\bk,t)=|{\bf E}({\bf k},t)|^2  P_D(\bk,t_0),
\ee
where, using the orthogonality of the eigenvectors, it follows
that
\be
\label{A}
|{\bf E}({\bf k},t)|^2=\sum_{n=1}^3\left[U_n(\bk,t)+\frac{2}{3 t_0}V_n(\bk,t)\right]^2(\hat{\mathbf{e}}_n\cdot\hat\bk)^2\,.
\ee
It is simple to verify \cite{marcos_06} that the ensemble
average of $P_D(\bk,t)/N$ over realizations of the IC 
is equal to the Fourier transform, as defined 
in  Eq.~(\ref{def-discreteFT}), of the displacement-displacement 
correlation function  $\langle \bu (0, t) \cdot \bu (\bR, t) \rangle$
(where $\langle..\rangle$ denotes the ensemble average).
 
To quantify the effects of  discreteness, it is convenient to define
the normalized quantity: 
\bea
\label{dev_ampl}
D_{\rm disp}(\bk,t)
&\equiv& \frac{P_D(\bk,t)}{P_D^{\rm fluid}(k,t)}
\nonumber\\
&=&\frac{|{\bf E} (\bk,t)|^2}{|{\bf E}^{\rm fluid}({\bf \hat k},t)|^2}
\nonumber\\
&=&\frac{|{\bf E} (\bk,t)|^2}{a^2(t)}
\eea 
where $P_D^{\rm fluid}(k,t)$ is the trace of the power spectrum 
of displacements evolved in the fluid limit.

\begin{figure}
\resizebox{8cm}{!}{\includegraphics*{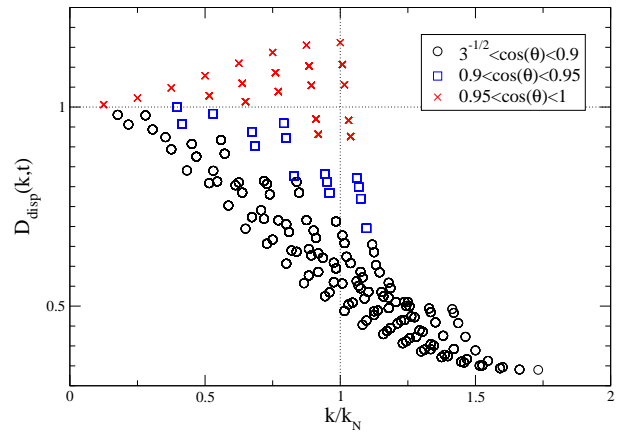}}
\caption{Discreteness factor $D_{\rm disp} (\bk,t)$ quantifying the
 modification with respect to the fluid limit ($D_{\rm disp} (\bk,t)=1$)
 of the power in the mode $\bk$ of the evolved displacement field.  The
 plot is given at at $a=5$ (for a simulation starting at $a=1$). See
 text for further details.
\label{figAD}}
\end{figure}

In Fig.~\ref{figAD} is shown $D_{\rm disp}(\bk,t)$
for the value $a=5$, i.e., after evolution from IC 
set at $a=1$. Deviations from unity are a direct 
measure of the modification of the theoretical (fluid) evolution 
introduced by the discreteness up to this time.
Note that $D_{\rm disp}(\bk,t)$ is plotted as a function of $k$,
each point corresponding to a different value of ${\bf k}$.  
The fact that the evolution depends on the orientation of 
the vector ${\bf k}$ is a manifestation of the breaking of 
rotational invariance by the lattice discretisation. The 
three different symbols for the points correspond to three 
different intervals of the cosine of the minimum angle 
$\theta$ between the vector ${\bf k}$ and one of the axes 
of the lattice. We will return to this point in further detail below.

\begin{figure}
\resizebox{8cm}{!}{\includegraphics*{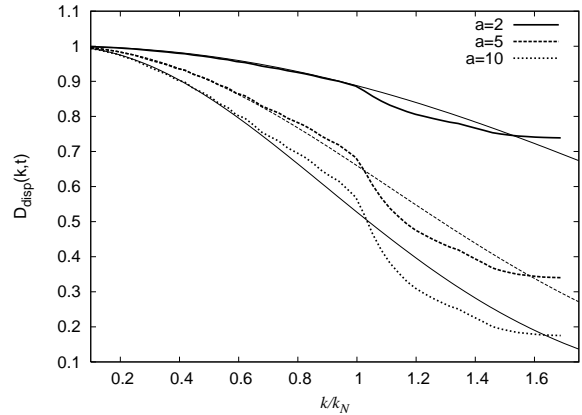}}
\caption{Discreteness factor $D_{\rm disp} (\bk,t)$ for $a=2$, $a=5$
and $a=10$, averaged over bins of $\Delta |\bk|=0.03 k_N$ for a $64^3$
simple cubic lattice. In each case is plotted also the equivalent
average of the analytic approximation given in
Eq.~(\ref{E-asymptotic-smallk}).
\label{fig-variousk-disp}}
\end{figure}

We have chosen the value $a=5$ because it is of the order of 
the typical one at which 
shell crossing occurs in $N$-body simulations, i.e., the initial 
amplitude in these simulations is typically such that this is 
the case\footnote{This
is true, for example, in the very widely used COSMICS 
package \cite{bertschingercode} for generating IC. The code 
in fact chooses the initial red-shift
given the number of particles, the comoving box-size and
the cosmological model, using the criterion that the
maximal mass density fluctuation of the 
realization on the grid is normalized to unity. This
gives a variance at the lattice spacing which is
less than, but of the order of, unity.}.
As we have already noted, the discrete evolution and
the fluid one are described by different exponents,
so this figure of $D_{\rm disp}(\bk,a)$ will change
as we change $a$. The resemblance between
Fig.~\ref{figAD} and the optical branch of the dispersion
relation in Fig.~\ref{fig1} allows us to infer simply this 
evolution: the reason for this resemblance is that 
already at the time chosen ($a=5$), the expression \eqref{A} 
is well approximated by the regime $t \gg t_0$,
given by Eq.~(\ref{E-asymptotic}), in which the most rapidly 
growing eigenmodes, which are those on the optical branches,
dominate. Using Eq.~(\ref{E-asymptotic}) it is easy to verify 
that
\bea
&&D_{\rm disp}(\bk,t \gg t_0)  \approx \nonumber \\
&&\left[\frac{{(\bf \hat k} \cdot {\bf{\hat e}}_{\rm max} ({\bf k}))(2+3\al_{\rm max}^+(\bk))}{3
(\alpha^-_{\rm max}(\bk)+ \alpha^+_{\rm max}(\bk))}\right]^2 
\left( \frac{t}{t_0}\right)^{2[\alpha^-_{\rm max}(\bk)-\frac{2}{3}]}\,. 
\label{E-asymptotic-D-disp}
\eea
For small $k$ (compared to $k_N$) we can simplify this expression.
In this case \cite{marcos_06} the eigenvalues on the optical
branch can be treated in a Taylor expansion about the fluid limit: 
\be
\label{expansion}
\ep(\bk)\approx  (1-b({\bhk}) k^2 \ell^2)\,, 
\ee 
where $b(\bhk)$ is a dimensionless coefficient
of order unity which depends on the direction in reciprocal space.
This expression is in fact \cite{marcos_06}, for the simple cubic
lattice, a good approximation to the eigenvalues up to 
$k \approx k_N$.  Using it we have 
\bea
\nonumber
 D_{\rm disp} (\bk, a) &\approx& 
\left(1+\frac{12}{25}b(\hat\bk)\ell^2 k^2\right) \\
&\times & ({\bf \hat k} \cdot {\bf{\hat e}}_{\rm max} ({\bf
k}))^2  a^{-\frac{6}{5} b(\hat{\bk}) k^2 \ell^2 }
\label{E-asymptotic-smallk}
\eea
and therefore
\be 
D_{\rm disp} (\bk, a
\gg 1) \approx ({\bf \hat k} \cdot {\bf{\hat e}}_{\rm max} ({\bf
k}))^2 \cdot a^{-\frac{6}{5} b(\hat{\bk}) k^2 \ell^2 } \,.
\label{E-asymptotic-largea}
\ee
In Fig.~\ref{fig-variousk-disp} is shown
$D_{\rm disp}(\bk,a)$, now averaged over all $\bk$ in a 
narrow bin of $k$, for different values of $a$. As anticipated
we see that the discreteness effects, i.e., the deviation from
unity of this quantity, grows as a function of time. Also
shown are the corresponding averages of the 
approximate expression Eq.~(\ref{E-asymptotic-smallk}) \footnote{We 
neglect here also, for simplicity, the time independent pre-factor. 
We will see below (cf. Eq.~\eqref{aniso-factor-asym} and 
Fig.~\ref{aniso-planewaves}) that it is indeed very close to unity
for $k < k_N$.}. We see that the agreement is very good, as expected, 
for sufficiently small $k$ and $a$.

\subsection{The power spectrum of density 
fluctuations}
\label{The power spectrum of density fluctuations}

We have considered above the evolution of the two point
properties of the displacement fields. Usually in the cosmological 
context one is more interested in a direct characterization of
the evolution of the density fluctuations, notably through the power
spectrum of density fluctuations or its Fourier transform, the 
reduced two-point (density-density) correlation function. 
In the cosmological literature the relation between the
two quantities is invariably derived in the continuous limit 
(see e.g. \cite{sb95}). Instead we use here the more
general result derived in \cite{andrea}, which gives
the PS of density fluctuations of a discrete
distribution of points generated by applying a generic
stochastic  displacement field with two point correlation matrix 
$S_{\mu \nu}(\br)$ to a distribution with initial PS 
of density fluctuations $P_{in}(\bk)$:
\begin{eqnarray}
\label{P_3d}
\nonumber P(\bk)&=& \int  d^3r
e^{-i\bk\cdot \br}
e^{-k_\mu k_\nu [S_{\mu\nu}(0)- S_{\mu\nu}(\br)]}
\left(1+\tilde\xi_{in}(\br)\right)\\ &-&(2\pi)^d\delta(\bk),
\end{eqnarray}
where $\tilde\xi_{in}(\br)$ is the Fourier transform of $P_{\rm in}
(\bk)$, i.e., the reduced two point correlation function of the initial
distribution, and the integral is over all space (as the expression 
applies in the infinite volume limit, i.e., to an infinite point 
distribution).

Expanding the exponential factor 
$e^{-k_\mu k_\nu [S_{\mu\nu}(0) - S_{\mu\nu}(\br)]}$
to linear order one obtains
\begin{eqnarray}
\label{PS_smallk} 
P(\bk)&=&P_{in}(\bk)+k_\mu k_\nu S_{\mu\nu}(\bk)  \\
\nonumber
&+&\frac{k_\mu k_\nu}{(2\pi)^3}\int d^3q 
S_{\mu\nu}(\bq) [P_{in}(\bk+\bq)
-P_{in}(\bk)]\,.
\end{eqnarray}
Note that in reciprocal space we now have an integral, rather
than a sum, as the density field is defined everywhere 
in real space. For the case we are studying here, however, of a 
lattice with finite periodicity, the displacement field inherits
the same periodicity and thus the function $S_{\mu\nu}(\bk)$ is 
non-zero only on the reciprocal lattice. The integral therefore
reduces again to a sum. 

In \cite{discreteness1_mjbm} we have 
studied in detail\footnote{See,
specifically, section IIIC and the appendices of \cite{discreteness1_mjbm}.}
the domain of validity of this linearized approximation 
Eq.~\eqref{PS_smallk} to the full
expression Eq.~(\ref{P_3d}). This depends in general on the shape
of the input PS $P_{th}(k)$, but for the PS in the range typical of 
cosmological models the criterion is just that the variance 
of the relative displacement of particles be small compared 
to the distance separating them. This coincides precisely
with the range of validity of PLT. Using then Eq.~(\ref{PS_smallk})
with the PS of the displacements as given by Eq.~(\ref{evolution-Pdisp}), 
we obtain
\begin{eqnarray}
\label{PS-density-evolved}
&&P(\bk,t)=P_{in}(\bk)+ ({\bf \hat k}\cdot {\bf E}(\bk, t))^2 P_{\rm th} (k)  \\
\nonumber
&&+\frac{k^2}{(2\pi)^3}\int d^3q 
({\bf \hat k}\cdot {\bf E}(\bq, t))^2 \frac{P_{\rm th} (q)}{q^2} [P_{in}(\bk+\bq)
-P_{in}(\bk)]
\end{eqnarray}
where we have used the Poisson equation with Eqs.~(\ref{zeldo-rel})
to relate the input theoretical PS $P_{th} (k)$ to the applied
displacement fields through
\begin{equation}
\label{theor-disp}
k^2 |{\tilde u}({\bf k},t_0)|^2= P_{\rm th} (k, t_0)\,.
\end{equation}

Eq.~(\ref{PS-density-evolved}) is thus the expression for the  
evolved PS of the density fluctuations in and $N$-body simulation
(on a perturbed lattice) as given by PLT.  The effects of 
discreteness described 
may be divided into two: 
\begin{itemize}
\item 
Effects encoded in the dependence of the function ${\bf E} (\bk ,t)$
on the lattice spacing $\ell$, which give as we have discussed 
deviations from its fluid value ${\bf \hat k} a(t)$.
This is manifestly a {\it dynamical} effect of discretization
on the lattice.

\item 
Effects encoded in the {\it additional} dependence on
the lattice spacing $\ell$ of the first and third term on 
the right hand side of Eq.~(\ref{PS-density-evolved}). 
This dependence comes through the initial PS $P_{in}(\bk)$.
These terms describe the density fluctuations which are
associated to the discrete sampling of the PS given by the 
second term on the right hand side of Eq.~(\ref{PS-density-evolved}),
on the lattice. They thus correspond to a {\it static} effect of 
the discretization. 
 
\end{itemize}

The first term on the right hand side of
Eq.~(\ref{PS-density-evolved}) depends only on the second effect ---
and is thus independent of time, describing simply the static
contribution of the initial unperturbed lattice --- while the second
term depends only on the first effect.  The third term, however,
couples both effects. It describes, as we shall see, how purely
discrete power, at large $k$ is induced and evolves in time.

$P_{in}(\bk)$, which is the PS of the lattice, is simply proportional 
to an infinite sum of delta functions at all points of the reciprocal 
lattice, i.e., at $\bk=2k_N {\bf m}$ where ${\bf m}$ is any vector of
non-zero integers. This term thus vanishes in the first Brillouin
zone [as defined after Eq.~(\ref{def-epsilon}) above]. It is 
straightforward to show \cite{discreteness1_mjbm} that the same 
is true for the third (convolution) term in 
Eq.~(\ref{PS-density-evolved}), provided the input PS is 
appropriately set equal to zero outside the first Brillouin 
zone\footnote{When applying the displacement field  using
Eq.~(\ref{theor-disp}) one can consistently include 
modes at larger wavenumbers. However this leads to aliasing
effects in the PS described precisely by this convolution
term (i.e. to the appearance of unwanted power at
long wavelengths). See \cite{discreteness1_mjbm} for further
detail.}. The second term on the right hand side of 
Eq.~(\ref{PS-density-evolved}) is thus the only term which is 
non-zero inside the first Brillouin zone\footnote{As discussed
in \cite{discreteness1_mjbm}, this is not true for a glass 
pre-initial configuration. In this case the third term 
in Eq.~(\ref{PS-density-evolved}) contributes at all $k$, 
and is proportional to $k^2$ at small $k$. See discussion in
conclusions section.}. The non-trivial discreteness effects
described by the second and third terms 
on the right hand side of Eq.~(\ref{PS-density-evolved}) 
thus contribute in non-overlapping regions of reciprocal
space.

\subsubsection{PS inside first Brillouin zone}

As we have just seen the evolved PS inside the first Brillouin 
zone, to linear order in the input PS, is
\be
P(\bk,t) = ({\bf \hat k}\cdot {\bf E}(\bk, t))^2 P_{\rm th} (k)\,. 
\label{approx-ps}
\ee
To characterize the effects of discreteness we thus define 
\bea
\label{dev_ampl}
D_{\delta \rho}(\bk,t)
&\equiv& \frac{P(\bk,t)}{P^{\rm fluid}(k,t)}
\nonumber\\
&= &\frac{|{\bf \hat k} \cdot {\bf E} (\bk,t)|^2}{|{\bf \hat k} \cdot {\bf E}^{\rm fluid}({\bf \hat k},t)|^2}
\nonumber\\
&=&\frac{|{\bf \hat k} \cdot {\bf E} (\bk,t)|^2}{a^2(t)} \,,
\eea 
where, using the definitions given above, we have
\be
|{\bf \hat k} \cdot {\bf E}(\bk,t)|^2= 
\left [ \sum_{n=1}^3 [U_n(\bk,t)+\frac{2}{3 t_0}V_n(\bk,t)](\hat{\mathbf{e}}_n\cdot\hat\bk)^2 \right]^2
\label{Efactor-density}
\ee
instead of the expression in Eq.~(\ref{A}) for the
analogous quantity for the PS of the displacement fields.

\begin{figure}
\resizebox{8cm}{!}{\includegraphics*{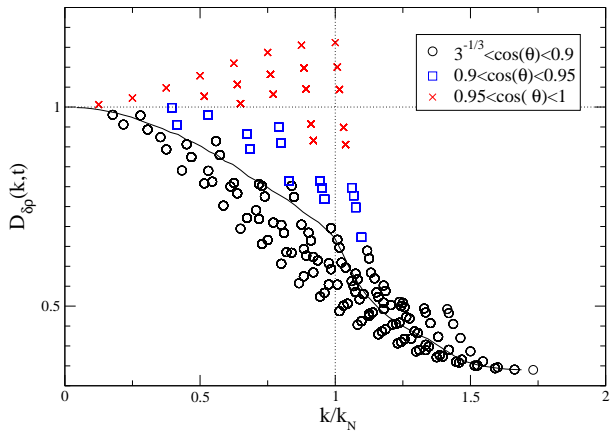}}
\caption{Discreteness factor $D_{\de\rho} (\bk,t)$ quantifying the
modification with respect to the fluid limit ($D_{\de\rho} (\bk,t)=1$)
of the power in the mode $\bk$ of the evolved density fluctuation field,
for $\bk$ in the FBZ.  The plot is given, as for Fig.~\ref{figAD}, at
$a=5$ (for a simulation starting at $a=1$), for a $16^3$ simple cubic
lattice. Also shown is a curve corresponding to the same quantity
averaged over all $\bk$ with $k$ in one of bins of equal width $\Delta
k=0.03 k_N$, for a $64^3$ simple cubic lattice. (The points of this 
$64^3$ calculation, which are not shown, just trace more densely
the behaviour of the points shown.)
\label{fig-discreteness-PS}}
\end{figure}

In Fig.~\ref{fig-discreteness-PS} is shown $D_{\delta \rho}(\bk,t)$, 
at $a=5$ as for the analogous Fig.~\ref{figAD} for the displacement 
fields in the previous subsection. The two figures are in fact very
similar, particularly at smaller $k$. The reason is the same one which
explained the similarity between  Fig.~\ref{figAD} and the optical
branch of Fig.~\ref{fig1}: the most rapidly growing mode on this
branch already dominates at this time so that the difference between
the expression in Eqs.~(\ref{A}) and (\ref{Efactor-density}) reduces
to a trivial time independent factor. Indeed we have now
\bea
&&D_{\rm \delta \rho }(\bk,t \gg t_0)  \approx {(\bf \hat k} \cdot {\bf{\hat e}}_{\rm max} ({\bf k}))^4 \\
&&\times\left[\frac{(2+3\al_{\rm max}^+(\bk))}{3
(\alpha^-_{\rm max}(\bk)+ \alpha^+_{\rm max}(\bk))}\right]^2 
\left( \frac{t}{t_0}\right)^{2[\alpha^-_{\rm max}(\bk)-\frac{2}{3}]}, \nonumber
\label{E-asymptotic-D-deltarho}
\eea
which differs from Eqs.~(\ref{E-asymptotic-D-disp}) 
only by the power of the product 
${\bf \hat k} \cdot {\bf \hat e}_{\rm max} (\bk)$.
We do not plot the analogous curves to those of
Fig.~\ref{fig-variousk-disp} as the results look
essentially the same.
 
At $a=5$ (i.e. at the time of shell crossing in a typical
cosmological $N$-body simulation) our Fig.~\ref{fig-discreteness-PS}
is a plot of the fractional discrepancy between the theoretically
evolved power (by FLT) and the power as evolved in the discretization
of this system (by  PLT). The fractional error introduced by
the discretization is largest, unsurprisingly, for the 
modes at the very largest wavenumbers ($k=\sqrt{3} k_N$, at the 
extremities of the first Brillouin zone), and decreases as 
$k$ does. At $a=5$ the power in the largest mode is reduced
to about {\it one third} of its fluid value, while around
$k=k_N$ the fractional error varies from about $+10 \%$ to 
more than $-50 \%$. At $k=k_N/2$ it varies from $+5 \%$ to 
about $-20 \%$, while at $k=k_N/4$ the total spread is about 
$10\%$. 

In Fig.~\ref{fig-discreteness-PS} is shown
also an average of $D_{\delta \rho}(\bk,a)$ (at $a=5$) 
over narrow bins of equal width in $k$, for a larger 
$64^3$ lattice. We see that this average is
dominated, at this time, by the more numerous modes with 
growth coefficients which are smaller than the fluid one.
We see that this average describes at all $k$ a 
{\it net slowing down} of the evolution of the power in the density 
fluctuations, ranging from slightly more than $30 \%$ at the 
Nyquist frequency, to $10 \%$ at half this frequency, and
down to about $3-4 \%$ at one quarter. 
It is straightforward to refine these estimates 
given the precise parameters of a simulation (i.e. the 
initial amplitudes to determine time of shell-crossing).
In our conclusions we will discuss the importance of 
these effects, and how they might be corrected for in 
simulations.

\subsubsection{PS outside the first Brillouin zone}

For $\bk$ outside the FBZ, we have
in Eq.~(\ref{PS-density-evolved}) only the non-trivial
contribution from the third term. Taking an 
input theoretical PS of the form  
\be
\label{PS-generalIC}
P_{th}(k) = Ak^n f(k/k_c)\,
\ee
where $f(k/k_c)$ is a function which cuts off the PS
at $k > k_c$,  it is straightforward to show (see
\cite{discreteness1_mjbm} for further detail) that
this term can be written
\be
\label{PS_order1_disc_simplified}
P_{\rm d}^{(1)}(k) = Ak^2k_N^{n-2} I(\mathbf{k}) 
\ee
where 
\begin{eqnarray}
I(\mathbf{k})= 
\sum_{\mathbf h\ne\mathbf 0} 
[\mathbf{\hat k}\cdot {\bf E}(\mathbf{h}-\mathbf{k},t)]^2
\left(\frac{|\mathbf{h}-\mathbf{k}|}{k_N}\right)^{n-2} \nonumber \\
\times f\left(\frac{|\mathbf{h}-\mathbf{k}|}{k_c}\right) 
\Theta_{FBZ}(\mathbf{h}-\mathbf{k}) 
\label{PS_order1_disc_simplified_sum}
\end{eqnarray}
where $\mathbf{h}=2k_N {\bf m}$, and the sum runs over all integer vectors 
${\bf m}$, and $\Theta_{FBZ}(\mathbf{h}-\mathbf{k})$ is a three dimensional
Heaviside function which is equal to unity inside the first Brillouin zone,
and zero elsewhere. This cut-off is imposed, as discussed above, 
in order to avoid aliasing effects.

At a given $\bk$ the sum $I(\mathbf{k})$ can pick up contributions 
only from a single vector $\mathbf{h}$, the one which lies inside 
the FBZ when translated by $-\bk$. It thus has the intrinsic 
periodicity of the lattice PS itself, i.e., 
$I(\mathbf{k})=I(\mathbf{k}+\mathbf{h})$, and it thus suffices 
to calculate it for vectors $\bk$ such that $k_N \leq k_i <2k_N$.
Just outside the FBZ it picks up contributions from ${\bf h}$ 
such that $\mathbf{h}-\mathbf{k}$ lies just inside the FBZ,
and thus of order $f(k_N/k_c)$, while at the larger values
$k \sim 2 k_N$ it picks up contributions from ${\bf h}$ 
such that $\mathbf{h}-\mathbf{k} \rightarrow 0$, with an
amplitude which depends strongly on the exponent $n$
of the input power spectrum. For any $n <2$, as is always
the case in cosmological models, the amplitude diverges
as one approaches $\bk = 2k_n {\bf m}$ (where $P_{in}(k)$
is also divergent). 

The term in the PS given in Eq.~(\ref{PS_order1_disc_simplified})
thus describes how the power {\it at all wavenumbers} in the 
input PS gives rise to power at small scales. Like the
power $P_{in}(\bk)$ associated with the unperturbed lattice,
it is {\it purely discrete}. Differently from this latter
term, which is time independent, it evolves in time as
described by PLT. It thus describes evolving physical clustering 
in the simulations {\it which is entirely
an artifact of discreteness}, rather than, as in the
effects analyzed above, a {\it modification} due to discreteness
of the non-trivial clustering of the fluid limit. Indeed it is 
interesting to note
that, even if we take $k_c << k_N$, i.e., a cut-off in the
input PS (as for example in typical in hot or warm dark matter 
simulations), 
this term is non-zero, and the growth can be well 
approximated by using the fluid limit for
${\bf E} (\bk, t)$ (since for the terms contributing in 
the sum $|\mathbf{h}-\mathbf{k}| \ell \ll 1$).
We will discuss this term further in our conclusions below.
   
\subsection{Measures of anisotropy}
\label{anisotropy}

\begin{figure}
\resizebox{8cm}{!}{\includegraphics*{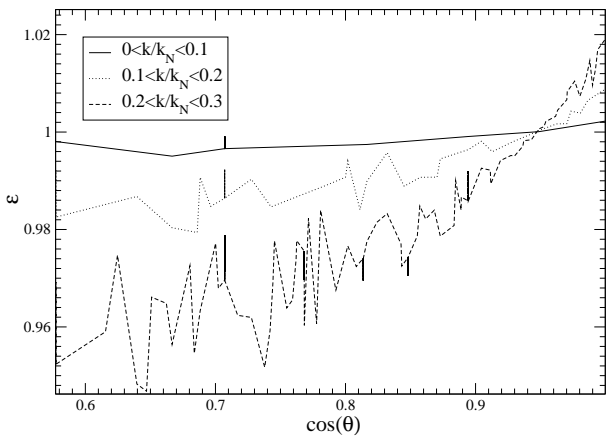}}
\caption{Average value $\varepsilon$  of
the normalized eigenvalues $\epsilon_n (\bk)=-\omega_n^2 ( \bk)/ 4\pi
G \rho_0$, averaged over the maximal eigenvalues at each $\bk$ for $k$
in the different ranges indicated, as a function of $\cos \theta$,
the minimal angle between $\bk$ and any axis of the lattice.
\label{aniso}}
\end{figure}

Let us consider in more detail the effects of anisotropy in the
evolution induced by the lattice discretization. The different
symbols indicated in Fig.~\ref{figAD} and 
Fig.~\ref{fig-discreteness-PS} correspond, as has been noted,
to the different ranges of the cosine of the 
minimum angle $\theta$ between the vector $\bk$ and 
one of the lattice directions. This is, of course, just a direct 
result of the dependence of the eigenvalues on this quantity evident 
in Fig.~\ref{fig1}, which is a manifestation of the 
anisotropy of the lattice. In Fig.~\ref{aniso} we plot the
eigenvalue as a function of $\cos \theta$
for the maximal eigenvalue (i.e. on the optical branch) averaged
over different ranges of $k$. The degree of anisotropy over the range 
of orientations is roughly characterized by the slope of these 
curves (as zero slope corresponds to isotropy). We see clearly
that the amplitude of the anisotropy thus decreases as the 
typical $k$ in the chosen bin decreases, which is as expected
as we approach the fluid limit [for which $\varepsilon=1$].
We observe also a clear trend in the deviation from the fluid eigenvalue:
this deviation typically increases as the orientation of the 
corresponding eigenvector goes further away from the axes of 
the lattice. This behaviour
is, however, not exactly followed, in particular around
$\cos \theta=1$ . Indeed it is notable that the eigenmodes 
oriented parallel, or very close to parallel, to the lattice
planes have eigenvalues which are larger than the fluid one.
We note that the very largest eigenvalue in Fig.~\ref{fig1},
associated with the largest values of $D_{\rm disp}$ and
$D_{\delta \rho}$ in Figs.~\ref{figAD} and \ref{fig-discreteness-PS}, 
correspond to an exactly longitudinal mode with $k=k_N$ and ${\bf k}$ 
parallel to the axes of the lattice. It thus describes the motion of pairs of 
adjacent infinite planes towards one another\footnote{Note that
this is true for the case only of a lattice with $N$ even.}.
The eigenvalue of this mode (see Fig.~\ref{fig1}) is 
$\epsilon \approx 1.1$, which gives [using 
Eqs.~\eqref{E-asymptotic-D-deltarho} 
and \eqref{alpha-epsilon}] 
$D_{\delta\rho} \propto a^{0.12}$. 

\begin{figure}
\resizebox{8cm}{!}{\includegraphics*{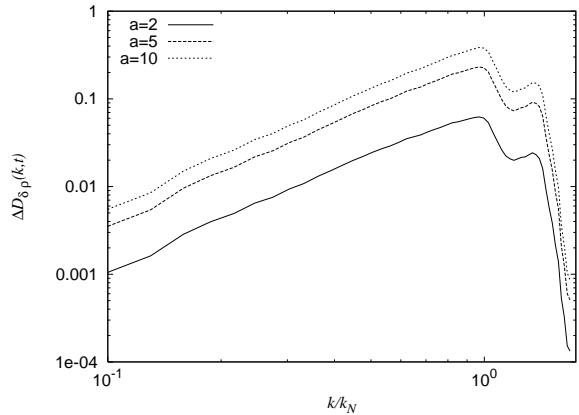}}
\caption{Discreteness factor $\Delta D_{\delta \rho} (k,t)$ 
for $a=2$, $a=5$ and
$a=10$, averaged over bins of $\Delta |\bk|=0.03 k_N$ for a
$N=64^3$ simple cubic lattice.
 \label{fig-aniso-dispersion}}
\end{figure}

There are clearly many different ways of quantifying these
effects. One simple measure is the dispersion in the growth
of the power in modes of the displacement or density fields.
Considering the latter we define 
\bea
\label{ddrho}
\Delta D_{\delta \rho}(k,t) &=& 
\left(\frac{\overline{D_{\delta \rho}^2} (k,t) - 
\overline{D_{\delta \rho}}^2 (k,t)}
{\overline{D_{\delta \rho}}^2 (k,t)}
\right)^{1/2} \\
&=& \nonumber\left[
\overline{\left(\frac{|{\bf \hat k} \cdot {\bf E}(\bk, t)|^2 -
{\overline{|{\bf \hat k} \cdot {\bf E}(\bk, t)|^2}}}
{{\overline{| {\bf \hat k} \cdot {\bf E}(\bk, t)|^2}}}\right)^2}\right]^{1/2} \,,
\eea
where the second equality is valid in the FBZ.
The average, of any function ${X}(\bk, t)$ on the reciprocal
lattice, is defined as  
\be
\overline{X} (k,t) = \frac{1}{N_k} 
\sum_{\bk , |\bk|=k} 
{X}(\bk, t)\,, 
\ee
$N_k$ being the number of eigenmodes at a given $k$. 
$\Delta D_{\delta \rho}(k,t)$ is evidently defined so that
it is zero in the  absence of anisotropy, and in particular in the fluid limit.
Its value, averaged in narrow bins of $k$, is plotted
in Fig.~\ref{fig-aniso-dispersion}, for three different
times corresponding to the scale-factors $a$ indicated
(where, as before, $a=1$ corresponds to the beginning 
of the simulation). The
results are very much as one would anticipate from
Fig.~\ref{fig-discreteness-PS}: the anisotropy is greatest
around $k=k_N$ where the spread in the growth factors is
greatest, and the amplitude at any $k$ grows as a 
function of $a$. Note that for a typical $N$-body simulation,
with $a=5$ at shell crossing, the dispersion in the growth
factor at this time at $k=k_N$ is of order $20\%$.

\begin{figure}
\resizebox{8cm}{!}{\includegraphics*{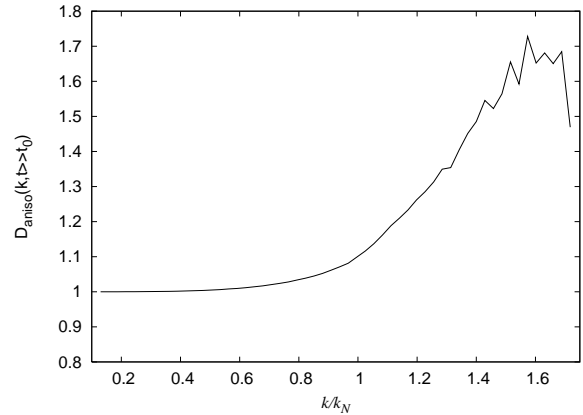}}
\caption{The asymptotic time independent anisotropy factor 
$D_{\rm aniso} (k,t>>t_0)$ defined in the text
averaged over bins of $\Delta |\bk|=0.03 k_N$, for a $64^3$ simple
cubic lattice.
\label{aniso-planewaves}}
\end{figure}

One simple manifestation of the anisotropy introduced 
in the evolution by the discretisation on a lattice 
was noted by Melott et al. in \cite{melott_all} :
an initial perturbation described by a pure plane wave, which 
in the fluid limit should maintain its purely one dimensional 
character, can become three dimensional. The breaking of symmetry
associated is, up to numerical precision, entirely associated to 
the lattice. In \cite{melott_all} the authors studied
this effect numerically, notably as function of the smoothing
at small scales. One simple quantitative characterization of
this effect is  
\be
\label{daniso}
{D}_{\rm aniso} (\bk, t)= \frac
{|\bf {\tilde u} (\bk,t)- {\bf \hat k} [{\bf \hat k} \cdot \bf {\tilde u} (\bk,t)]|^2}
{|{\bf \hat k} \cdot \bf {\tilde u} (\bk,t)|^2},
\ee
i.e., the fractional deviation of the 
displacement\footnote{In \cite{melott_all} the analogous quantity
is considered, but for the  velocity rather than the displacement. 
It is straightforward also to calculate this quantity; we 
choose the displacement as it makes the calculation a little
simpler.}
direction parallel to $\bk$. It follows from 
Eq.~(\ref{eigen_evol_cosm}) that 
\be
{D}_{\rm aniso} (\bk, t)= \frac
{{\bf E}^2 (\bk,t)}{({\bf \hat k} \cdot {\bf E}(\bk,t))^2}\,,
\label{aniso-factor}
\ee
which is evidently unity in the fluid limit. For
$t \gg t_0$ it follows, from Eqs.~(\ref{E-asymptotic-D-disp}) and
(\ref{E-asymptotic-D-deltarho}), that
\be
{D}_{\rm aniso} (\bk, t \gg t_0)= \frac
{1}{({\bf \hat k} \cdot {\bf e}_{\rm max} (\bk))^2}\,,
\label{aniso-factor-asym}
\ee 
i.e., it is independent
of time, depending only on the orientation of the eigenvector of the
maximally growing mode with respect to $\bk$. This quantity, averaged
over narrow bins of $k$, is shown in Fig.~\ref{aniso-planewaves}. 
We see the convergence to the fluid limit, 
in which ${\bf \hat e}_{\rm max}(\bk) \rightarrow {\bf \hat k}$, 
for $k/k_N \ll 1$ and increasing deviations as $k$ increases.

\section{Discreteness effects with a small scale smoothing}

In cosmological $N$-body simulations a smoothing at small scales
of the interparticle (Newtonian) potential is generally
employed \cite{efstathiou_init, couchman}. ``Small'' in this context 
means that the 
characteristic scale for this modification of the potential
is smaller or comparable to the average initial inter-particle 
distance, i.e., approximately the lattice spacing $\ell$ in the 
case we are considering, of perturbations from a lattice.
Almost always in fact this smoothing is taken to be
considerably (i.e. one to two orders of magnitude)
smaller than $\ell$; an exception is in $PM$ simulations 
in which the effective smoothing scale is normally
equal to the scale $\ell$, as the potential is calculated 
on a mesh which typically is taken to coincide with the 
initial lattice  configuration. There is in this case no correction to take 
into account forces from sub-mesh particles, as in $P^3M$ or 
tree codes (see e.g. \cite{efstathiou_init, couchman}).

The motivation for the use of such a smoothing in this
context is that it should reduce the 
effects of discreteness and make the simulation approximate better
the theoretical (fluid-like) behaviour. The reason for this
expectation is that it reduces the impact of hard two-body collisions,
which should be negligible in the fluid limit
(in which the mean-field dominates). However,
there is no demonstration (more than these approximate
qualitative arguments) that such a smoothing has the
effect of making a simulation a better approximation
to the desired fluid limit. Indeed attempts in
the literature to treat this question numerically
(e.g. \cite{splinter, kuhlman, melott_alone, 
melott_all, diemandetal_convergence, diemandetal_2body, 
binney_discreteness, discreteness-hamana})
lead to quite different conclusions.
Notably the use of a smoothing smaller than $\ell$ has
been contested by some authors 
\cite{splinter, kuhlman, melott_alone, melott_all}, 
while such a practice is almost universal in large 
current $N$-body simulations.

Let us consider what we can learn about this question
using the methods developed in this paper. As was 
noted in Sect.~\ref{PLT}, the perturbative analysis
we have described can be applied to any two-body interaction
potential, and in particular to the case of a smoothed Newtonian
potential. We can simply repeat our calculation of 
the eigenvalues and eigenmodes given the potential.
The form of the smoothing employed varies and has evolved 
in time. Currently it is common to use a smoothing which leaves 
an exact $1/r$ potential for $ r > \epsilon$, where $\epsilon$ 
is the scale characteristic of the smoothing. If 
then $\epsilon \ll \ell$, as is typically the case, this 
modification has no effect until two particles are closer 
than $\epsilon$. It therefore has no effect on our 
calculation (which breaks down when this condition is 
fulfilled anyway). Thus such a smoothing has no effect 
in moderating or otherwise the discreteness effects we 
have quantified. The fact that that this is the case 
illustrates clearly an important point: the discreteness 
effects we have described are {\it not two-body collisional 
effects}. 

In order for a smoothing of the potential to be of relevance 
to the effects we are describing it must modify the 
potential at the interparticle distance, as we do a
perturbation expansion about the configuration with
the particles at this distance. Let us consider therefore 
a simple smoothing which does so, the so-called 
``Plummer'' smoothing: 
\be
\label{smoo}
\phi(r)=\frac{1}{\sqrt{r^2+\epsilon^2}}.  
\ee 
When $\epsilon \sim \ell$ it
would be expected to, and indeed does, produce results
similar to $\mathrm PM$ simulations (see e.g. \cite{splinter, melott_alone}) 

\begin{figure}
\resizebox{8cm}{!}{\includegraphics*{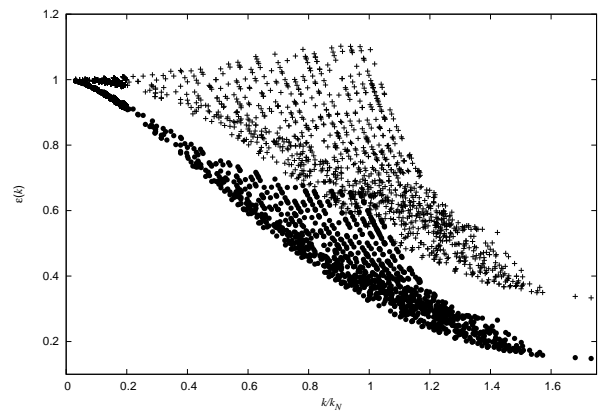}}
\caption{Eigenvalues of the eigenmodes of the displacement fields
for full gravity (top set), and  with a smoothing at half the 
interparticle distance (see text), for a $64^3$ simple cubic
lattice. For clarity only $0.5\%$ of the eigenvalues are shown for 
$k/k_N>0.2$.
\label{e_smoo}}
\end{figure}

For a given value of $\epsilon$ we can calculate the
spectrum of eigenvalues and eigenmodes. 
As one would anticipate, the modifications to them with 
respect to the case of pure gravity, are very small 
for $\epsilon \ll \ell$, and  
only becomes significant when we have $\epsilon \sim \ell$.
In Fig.~\ref{e_smoo} we show the eigenvalues on
the optical branch, for  $\epsilon=\ell/2$. Also plotted, for
comparison, is the same branch for pure gravity.

The conclusions which follow from this figure are very
clear. The effects of introducing a significant smoothing is
to {\it increase the average amplitude of discreteness effects, 
while decreasing the effects of anisotropy}. These behaviours 
are straightforward to understand qualitatively. The contribution 
to the total gravitational forces coming from nearby particles starts 
to be reduced significantly when the smoothing becomes of order
the interparticle distance. Thus the growth factors at these
scales decrease relative to the case of full gravity, but 
they also show less anisotropy as the strong anisotropy in the
distribution of nearest neighbors is no longer felt
as much.

\begin{figure}
\resizebox{8cm}{!}{\includegraphics*{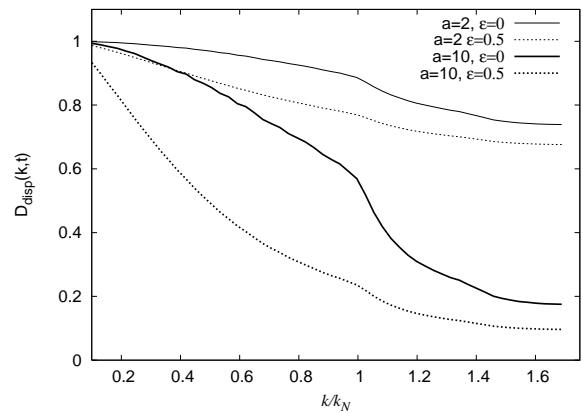}}
\caption{Discreteness factor $D_{\rm disp} (\bk,t)$ for two different
values of the smoothing parameter $\ep$ (in units of $\ell$), for
$a=2$ and $a=10$ for a $64^3$ simple cubic lattice.  An average over
bins of width $\Delta |\bk|=0.03 k_N$ has been performed.
\label{smoothing-D-deltarho}}
\end{figure}

\begin{figure}
\resizebox{8cm}{!}{\includegraphics*{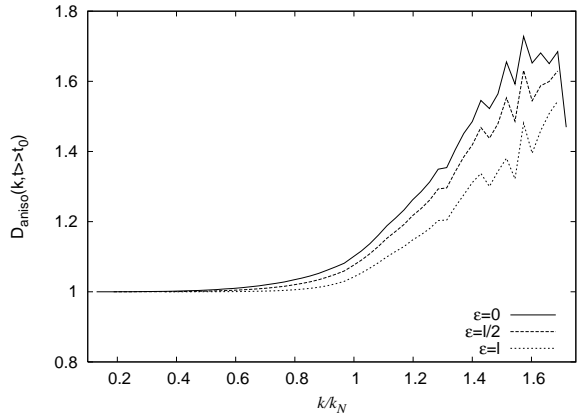}}
\caption{The asymptotic anisotropy factor $D_{\rm aniso} (k,t\gg t_0)$ 
averaged over bins of $\Delta |\bk|=0.03 k_N$, for three values of
the smoothing parameter $\epsilon$, for a $64^3$ simple cubic lattice.
\label{aniso-planewaves-smoo}}
\end{figure}

These effects can easily be further quantified by considering the
effects of varying the smoothing scale $\epsilon$ on the quantities
we defined above. In Fig.~\ref{smoothing-D-deltarho} we show the 
fluid-normalized discreteness factor $D_{\rm disp}$ [as defined
in Eq.~\ref{dev_ampl}] both without
any smoothing (as in Fig.~\ref{fig-variousk-disp}) and 
for $\epsilon=\ell/2.$ The result is self-evident given the results
for the optical branch in Fig.~\ref{e_smoo} above: the smoothing
increases the effects of discreteness on average.
The expected reduction in anisotropy as $\epsilon$ increases is
illustrated in Fig.~\ref{aniso-planewaves-smoo}, which shows the asymptotic
value of the function ${D}_{\rm aniso}$ 
[as defined in Eq.~\ref{daniso}] for three values
of $\epsilon$. As the latter parameter increases, we see
that on average the optical modes become more longitudinal
which results in a reduced symmetry breaking effect in
the propagation of plane waves. This is
indeed the conclusion of the numerical study of this
discreteness effect in \cite{melott_all}. We note, however,
that the present analysis leads to a conclusion
different to that of \cite{melott_all}: while 
the smoothing reduces this effect of discreteness, it actually 
increases the averaged effect of discreteness. By increasing 
the smoothing scale one obtains a better approximation to fluid
behaviour only in the limited sense that one recovers better one 
qualitative feature of fluid behaviour, its isotropy. 

\section{Parametric and limiting behaviours}
\label{Parametric and limiting behaviours}

We have given in the precedent sections various measures of
the modification of the fluid limit evolution engendered
by discretization on a lattice. We have made use throughout
of the result of \cite{joyce_05, marcos_06}, which we
have summarized in Sect.~\ref{The Zeldovich Approximation},
that the fluid limit for the evolution {\it of a given mode} is
recovered exactly when we send $\ell \rightarrow 0$ 
while keeping $k$ fixed. We have also seen that, keeping $\ell$
fixed, the deviations of the evolution of this same mode from 
its fluid behaviour at shell crossing increases as the time
of shell crossing does. This implies that the evolution of
cosmological simulations deviates completely from the desired
theoretical evolution in the limit that the starting red-shift
is increased keeping all other parameters fixed. Given these 
subtleties about how the continuous limit is recovered, it 
is instructive to consider these points in greater detail.

\subsection{The fluid limit}

Let us consider carefully in what limit the desired fluid limit
for the evolution of a cosmological $N$-body simulation (in the
class considered here) is recovered. 

Such a simulation, periodic on a cube of side $L$, is 
a discretisation of the continuum problem, which introduces, 
at least, the following parameters:  
\begin{itemize}
\item the lattice spacing $\ell$, given by $\ell=L/ N^{1/3}$,
\item the force smoothing scale $\epsilon$, and 
\item the initial red-shift $z_{\rm init}$.
\end{itemize}
The interpretation of results of the simulation 
as physical is justified if they approximate well
those which would be obtained by simulations with 
suitable extrapolations of these parameters towards
the theoretical continuum limit\footnote{The results must 
of course also be, to a sufficient approximation, independent 
of the box size $L$. Finite size effects of this
type are not the subject of study here, as they are distinct from
discreteness effects. Indeed such effects are equally present in the 
fluid limit (treated analogously with periodic boundary conditions).}.
The importance of understanding this limit precisely is that
it explains how this extrapolation of the parameters should
be performed in practice. 

The input theoretical cosmological model, on the other hand, is 
characterized fully, for Gaussian initial fluctuations, by its PS. 
If the side $L$ of the box is specified in physical units, the 
applicability of the ZA for setting up the IC 
then fixes, for given $\ell$, a {\it minimal} value of the initial 
red-shift $z_{\rm init}$. This corresponds to an upper bound on the 
initial amplitude of the PS. The PS is, in general, a function which 
involves numerous physical scales characteristic of the cosmological model. 
It contains necessarily, however, at least one such scale: that
characterising decay at large $k$ of the PS. Indeed since the
one point variance of the density fluctuations, which is proportional
to the integral of the PS, is necessarily finite, one has that 
\begin{equation}
\lim_{k \rightarrow \infty} k^3P(k)=0\,.
\end{equation}
There is thus always a physical scale given by a wavenumber $k_c$, 
above which fluctuations of density decay faster than $k^{-3}$.
In simulations of the currently favored cold dark matter (CDM) 
models $\ell$ is always, because of numerical limitations, much 
larger than $k_c^{-1}$ (i.e. $k_c$ is larger than
the particle Nyquist frequency), and the displacement 
field generated is then cut-off in the first Brillouin 
zone rather than at $k_c$ (to avoid aliasing effects). 
In hot or warm dark matter (HDM or WDM) models, on the other hand,
the (approximately exponential) cut-off is typically
at a wavenumber considerably smaller than the Nyquist
frequency.

The {\it exact} fluid limit, in the regime of application of PLT,
is that in which the corrections due to discreteness vanish, i.e.,
all the terms in Eq.~(\ref{disc_cosm}) vanish exactly and 
the particles follow the exact FLT trajectories. It is clear
that, strictly, this occurs only if we impose an abrupt
cut-off $k_c$ in the input spectrum of modes, which remains 
fixed as the particle number increases, i.e., the exact fluid
limit is $k_c \ell \rightarrow 0$ at fixed $k_c$. We can,
in practice, identify this $k_c$ with the (not necessarily
abrupt) cut-off in the input PS: modes for $k>k_c$ do
not contribute significantly to the usually physical 
relevant quantities (such as mass variance).

To specify fully the limit we need to prescribe how the
other parameters, $z_{\rm init}$ and $\epsilon$, are treated.
It is clear that it is sufficient to keep $z_{\rm init}$
fixed for what has been said above to hold. As for 
$\epsilon$, we have seen that it leads to a modification
of the growth of modes with respect to their fluid 
evolution, so the latter will only be recovered if
$\epsilon \rightarrow 0$ as $\ell \rightarrow 0$,
i.e., $\epsilon \rightarrow 0$ in physical units. 
Given that PLT is well defined for $\epsilon=0$ the
order in which one takes the two limits is clearly not
important.

\subsection{Long time/high initial red-shift limit}

We have seen that, because the modification of evolution 
associated to discreteness appears as a difference in the
exponents of growth of any given mode, the discrete
system, with a given $\ell$, diverges from the fluid one 
when one considers the long time limit for any given 
$k$ if {\it $\ell$ remains fixed}. Indeed for $k < k_N$ we 
have seen, for example, that the statistical measures of discreteness 
effects in the evolution of the PS of density fluctuations
$D_{\delta \rho}$ is well approximated by 
Eq.~\eqref{E-asymptotic-D-deltarho}. This expression
diverges at large times arbitrarily far from the 
fluid limit, $D_{\delta \rho}=1$.
These PLT approximations remain good until close to shell 
crossing at the scale $\ell$ which occurs at a red-shift 
determined by the cosmological model (and independent of 
the choice of the initial red-shift $z_{\rm init}$ for the 
simulation).
{\it Thus, for a given physical scale, discreteness effects 
increase without limit as the starting red-shift of the simulation 
is increased, at fixed particle number}. 
For asymptotically large values any physical quantity will
be dominated by the modes with the largest exponents.
In the case of the simple cubic lattice, as we have noted
above, these modes correspond to the collapse of infinite 
planes along the axes of the lattice.

It follows from this discussion that, in the fluid limit 
we defined above, the order of the limits for $N$ (or $\ell$) 
and for the duration $t$ of the simulation (or $z_{\rm init}$), 
is important, i.e.,  
\begin{equation}
\lim_{z_{\rm init} \rightarrow \infty} \lim_{\ell \rightarrow 0} \neq 
\lim_{\ell \rightarrow 0} \lim_{z_{\rm init} \rightarrow \infty}\,,
\end{equation}
or, equivalently,
\begin{equation}
\lim_{t \rightarrow \infty} \lim_{N \rightarrow \infty} \neq 
\lim_{N \rightarrow \infty} \lim_{t \rightarrow \infty}\,.
\end{equation}
This result is in fact not surprising, at least for those familiar
with the use of the Vlasov limit to describe systems (e.g. plasmas)
with long range interactions. It is in fact well known 
(e.g. \cite{braun+hepp, spohn}) that there
is a non-commutativity of the long time and large number limits
in such systems. The Vlasov limit corresponds to taking the 
particle number to infinity before the time. The long time limit
of a system with fixed $N$ is a different physical limit
of the system's behaviour. Indeed a good example of this are 
precisely the two-body collisional effects usually considered
as sources of discreteness error in $N$-body simulation: these
become relevant \cite{binney}, in a uniform system with mean 
density $\rho_0$  on a time-scale $t_{2} \sim t_{dyn} N/\ln N$
where $t_{dyn}=1/\sqrt{4 \pi G \rho_0}$. Thus clearly if one
considers the limit $t \rightarrow \infty$ at finite $N$ one never
recovers the Vlasov behaviour. Toy models in one dimension
manifesting this kind of behaviour have recently been studied
in detail (see e.g. \cite{yamaguchi_etal_04}).

Practically this observation implies that at least one of the conditions 
for keeping discreteness effects under control in
an $N$-body simulation is a constraint on the
initial red-shift (i.e. given $\ell$, that $z_{\rm init}$ be 
less than some value).  We note that the $z_{\rm init}$ is not a 
parameter which has been considered in discussions
of discreteness effects in $N$-body simulations in the literature 
(e.g. \cite{melott_all, splinter, melott_alone, kuhlman, discreteness-hamana}).

\subsection{$N$ dependence of effects}

It is instructive also to consider more explicitly
the $N$ dependence of the effects we have described. A characteristic
time scale for these effects, somewhat analogous to the one just refered
to for two-body relaxation, can be derived by considering the 
evolution we have described here in static space (i.e. with $a(t)=1$). 
The only modification 
with respect to the EdS case we have treated here is in the mode 
functions, which become simply growing and decaying exponentials. 
Fluctuations
on all but the shortest scales have eigenvalues which
may be well approximated by Eq.~(\ref{expansion}), or
\begin{equation}
\om (\bk) \simeq  [1- \frac{b({\bhk})}{2}  k^2 \ell^2)] t_{dyn}^{-1}
\end{equation}
where $\om (\bk) = t_{dyn}^{-1}$ is the fluid behaviour
[and $b({\bhk})$ is a constant of order unity].
For any quantity dominated by modes around $\bk$ 
the deviation from fluid behaviour will become significant
when  
\begin{equation}
e^{ b({\bhk})k^2 \ell^2 t/2t_{dyn}} \sim 1\,.
\end{equation}

The characteristic time scale for the discreteness effects 
to dominate may thus be estimated as
\begin{equation}
\label{timescale}
t_{disc} (k) \sim \frac{t_{dyn}}{k^2 \ell^2} \propto N^{2/3}
\end{equation}
where the last proportionality is inferred for a given $k$
(fixed in units of the box size as the particle number $N$
increases). Note that here $N$ is the number of particles
sampling the scale $\sim k^{-1}$ rather than the total number
of particles in the system. Thus the time scale for the 
discreteness effects to dominate is a function of the scale
considered rather than a global timescale for the system.

\section{Discussion and conclusions}
\label{Discussion and conclusions}

We have described in this paper the application of a formalism
developed in \cite{joyce_05, marcos_06} to determine the errors
in $N$-body simulations at early times, i.e., up to shell crossing
when the formalism breaks down. We have given an explicit expression,
Eq.~(\ref{disc_cosm}), for the error in the trajectory of any particle,
assuming only that IC are set up in the standard
way using the ZA. We have then defined,
evaluated numerically for a typical red-shift of shell-crossing,
and given approximate analytic expressions for, functions which 
characterize these discreteness errors for any given mode of
the displacement or density field. We have also discussed the
effects of a smoothing of the force at small scale on these 
effects, as well as the precise limits in which the fluid
limit of the system's evolution is recovered. 

We now discuss both the immediate practical relevance of these 
results --- their application in quantitifying  
{\it a source} of systematic error in simulations ---
as well as their relevance to the broader question of
discreteness error in $N$-body simulation, for other
kinds of initial configurations and beyond shell crossing.

\subsection{Application of results}

We have seen that, at a red-shift of shell crossing typical
for a cosmological simulation (with IC using
e.g. the code of \cite{bertschingercode}), errors larger than
a few percent have accumulated for all wavenumbers larger 
than about a quarter of the Nyquist frequency. Such modes
are usually included in the IC in 
simulations of CDM type models, which typically cut
the input PS at the edge of the first Brillouin zone.
Indeed this means that for most of the modes included
errors are larger : the modes with $k > k_N/4$
represent more than $99\%$ of the sampled modes
\footnote{The fraction of modes with $k < k_0$
is, approximately, the volume of a sphere of radius 
$k_0$ divided by the volume of the FBZ.}, 
while those with $k > k_N/2 $,  which have errors 
exceeding $25\%$, about $93\%$.
Any physical quantities at the end of the simulation which 
depends on the power in these modes must inherit {\it at
least} this error. Such power is included
usually as one is indeed interested in such quantities.

What precisely the induced error is will depend of course on
the quantity considered, and we do not attempt here to describe
such effects exhaustively. The results given in this 
paper allow one to calculate them, and thus in principle
correct for them, in a very straightforward manner: it 
suffices to calculate the function $D_{\rm disp} (\bk, a_{\rm plt})$
as described 
in Sect.~\ref{Discreteness effects in the power spectrum of
displacements}.
We have introduced here the scale factor labelled $a_{\rm plt}$
to indicate that up to which PLT is a good approximation. 
As we have discussed, and shown in detail by numerical study in
\cite{marcos_06}, this corresponds approximately to shell crossing. 
Its precise definition is of course somewhat subjective as it 
involves specifying how strong a deviation defines it (and in 
what quantity the deviation is measured).  Here it can be taken 
either as a theoretically
calculated shell crossing at the scale $\ell$ or, better,
as determined a posteriori from the simulation (e.g. by matching the
evolution of the variance of displacements to that described by 
PLT, as reported in \cite{marcos_06}). 

The effects of discreteness up to shell crossing may thus be 
considered as equivalent to a $\bk$-dependent misrepresentation
of the input displacement field (at $a=1$), by a factor in amplitude
of $1/\sqrt{D_{\rm disp} (\bk, a_{\rm plt})}$. The effective initial 
amplitude is, for almost all modes, somewhat less than the theoretical 
amplitude.  This could, in principle, be ``corrected'' for 
by multiplying  each mode $\bk$ of the input displacement field,
at the chosen $z_{\rm init}$ (corresponding to $a=1$), by the factor 
$1/\sqrt{D_{\rm disp} (\bk, a_{\rm plt})}$. Such a correction
would ensure that the full linearized evolution PLT coincides,
at $a=a_{\rm plt}$, with the analogous (linearized) fluid result.

It is not clear, however, that such a correction of the initial
amplitude will in fact produce a more accurate result than simply 
starting the simulation at $a=a_{\rm plt}$, using the ZA to set up the
IC at this time (and possibly improved taking 
higher order corrections in Lagrangian perturbation theory into 
account). Applying the discreteness correction to the IC at
$a=1$ ($<a_{\rm plt}$) means that the linear evolution is offset 
from its theoretical evolution for  $a < a_{\rm plt}$. It is 
possible that the integrated non-linear effect of this offset 
will be greater than the error involved in using the ZA directly at 
$a=a_{\rm plt}$ when the linear approximation it is based on is
breaking down.  In principle there exists a starting 
red-shift which is optimal in the sense that it minimizes the
combined errors due to non-linearity (which decreases as 
$z_{\rm init}$ increases\footnote{For evaluation of the errors
in the fluid limit due to non-linearity, see \cite{scoccimarro_transients_98,
valageas_IC_03}.}) and errors due to discreteness 
(which increases, as we have seen, as $z_{\rm init}$ does).
We note in this respect that the perturbative formalism 
used here, which described the fully discrete problem, can
be extended beyond the linear order we have used. Such
a treatment should allow a determination of this optimal
red-shift.

These comments apply to IC as they are 
applied in CDM type simulations. In HDM or WDM simulations 
the physical cut-off in the initial PS is, as has been 
noted above,  typically such that $k_c << k_N$. Thus the modes 
of the displacement
field which have the greatest modification of their evolution due to
discreteness are not present. For
modest typical values of $z_{\rm init}$ the effects just 
discussed are therefore negligible. Above, however, we discussed 
only the effects of the modification
of the dynamical fluid evolution, and not the additional
effects which have been described in 
Sect.~\ref{The power spectrum of density fluctuations}, in the
PS of density fluctuations outside the FBZ. We have shown
in our discussion of these terms that this additional,
purely discrete, contribution in the PS is present even if the
input PS is limited to a region well inside the FBZ,
as in HDM or WDM simulations. In the approximation that
the small $k$ modes grow according to fluid theory (i.e. are
linearly amplified in proportion to the scale factor $a$ in 
the EdS model) it is simply an extra contribution in the initial PS
which is also linearly amplified. It describes the growth
of power at small scales due to the coupling between long 
wavelength power and that intrinsic in the initial
lattice. We note that this physical mechanism generating
``artificial'' discrete power at scales of order the
initial interparticle distance has been observed and
studied in numerical simulations of 
HDM/WDM \cite{gotz+sommerlarsen_WDM, wang+white_HDM}.
Our treatment gives an analytic description of this 
process (up to shell crossing), for the case of an initial
simple cubic lattice. 

In this latter context comparative studies
have been made of this initial configuration with ones
generated starting from the ``glass'' configurations sometimes
used as an alternative \cite{white_leshouches}. The 
formalism we have used in this paper can in principle be generalized to
treat such a starting configuration, and thus to compare
analytically the discreteness effects up to shell crossing
in the different cases.
There is, however, a considerable technical complication:
the Bloch theorem does not apply and the eigenvectors
of the displacement field about the (now approximate)
zero force configuration are no longer plane waves.
Thus the diagonalization problem cannot be simplified
in the same way, and a fully numerical solution would
be required to determine the $3N$ eigenmodes. We expect
to find only small quantitative differences: the 
origin of the discreteness effects we have quantified 
on a lattice arise essentially, as we have discussed just above, 
from the sampling of the continuous density field by 
particles. The magnitude of the effects will depend 
essentially on the average particle density, and not 
on the details of the arrangement of the points in the 
distribution. 

We note, on the other hand, that the treatment of the 
problem on other cubic lattices is straightforward.
In \cite{marcos07} the cases of a body-centered cubic
and face-centered cubic lattice are solved. We will
exploit this treatment of discreteness effects on
a range of different ``pre-initial'' lattices in
a forthcoming numerical study of discreteness effects
in the non-linear regime \cite{preIC_07}.

\subsection{Discreteness in the non-linear regime}

The perturbative treatment presented here can, as has been noted, be
extended to beyond linear order. While such an approach may lead 
to useful insights into the interplay of discreteness effects and
non-linearity, it can extend at most as far as shell crossing.
The current formalism therefore allows us to say nothing
quantitatively about discreteness effects in the fully non-linear
regime. It can however, as we now discuss, give us some qualitative 
insight which may be useful in attempts to understand and quantify 
such effects.

Discreteness effects, we recall, are all those effects which lead
to differences between the evolution of statistical quantities
in $N$-body simulations and that in the theoretical Vlasov-Poisson
limit. We have underlined that the effects studied here, which
are manifestly of this type, are different in nature to those 
usually emphasized in this context:
(i) two-body collisional effects (see e.g \cite{diemandetal_2body, 
binney_discreteness, Baertschiger:2002tk}), and (ii) the limitations 
imposed by the particle discretisation on the representation 
of the initial theoretical PS (described analytically in \cite{discreteness1_mjbm}).
The modifications of the continuum evolution we have described 
here depend explicitly on the dimensionless ratio $k \ell$, 
and decrease in amplitude as $k \ell$ decreases. They can 
thus be well described, and distinguished from these other effects,
as 
{\it dynamical sparse sampling effects}, i.e., 
they are modifications of the evolution of fluctuations which 
arise from the fact that a given scale is sampled more
sparsely by the ``macro-particles'' of an $N$-body simulation
than by the microscopic particles of a fluid.

Clearly one would expect that such a physical effect should 
be present also in the evolution after shell-crossing, in addition
to two-body collisional effects and any effects inherited from
the IC. The difficulty in assessing its importance,
at a given scale and time, is that it is not clear what parameter
should play the role of $\ell$ after shell crossing. The results 
of simulations are usually interpreted as physical down to scales 
$\sim \epsilon$, where $\epsilon \ll \ell$. This is typically of the 
order of the interparticle distance in the {\it densest} regions of 
the final configurations (at the centers of halos). This extrapolation 
would appear then to assume, roughly, that the role of $\ell$ in
the non-linear regime is played by the {\it minimal} interparticle distance
in the evolved simulation. This seems, given the physical nature
of the effect, an extremely optimistic assumption. One would
expect fluctuations at significantly larger scales in any less 
dense region (i.e. most of the
volume) to be subject to much greater discreteness error of this
type. In very low density regions, notably, it is clear that all
but the very longest wavelengths (compared to the locally very large
interparticle separation) will be very poorly approximated.

Another related point which emerges from our analysis of the 
evolution up to shell crossing is the subtlety involved in
defining the correct continuum limit. We saw, in particular, that
an extrapolation to high initial red-shift at constant particle
number (and force smoothing) leads to a divergence from
the desired limit. We note that this is an unexpected
behaviour: variation of this parameter has been considered
as relevant only to the accuracy with which the IC are
represented \cite{scoccimarro_transients_98, valageas_IC_03}, 
as the Zeldovich approximation becomes exact
only in the limit that $z_{\rm init} \rightarrow \infty$.
Our results indicate therefore that numerical tests which 
vary $N$ to test for discreteness effects should best be 
done at fixed $z_{\rm init}$. Further we have seen that the continuum
limit is attained strictly only in the limit that 
$k_c \ell \rightarrow 0$ where $k_c$ is the cut-off in
the input PS. While we have been able to calculate corrections
outside this regime (i.e., for CDM type simulations which, as 
noted above, are performed with $k_c \ell \gg 1$) it is 
not evident that it will be possible to do so in the fully
non-linear regime.  This suggests
that, to calculate robust discreteness errors in the 
latter regime, it may indeed be necessary  
to extrapolate numerically to a regime in which $\ell$
(as argued notably in \cite{melott_all, splinter, melott_alone,
kuhlman}) 
--- or whatever length scale plays its role in the non-linear regime ---
is much less than all physical scales at which clustering is to be
determined. A related discussion of some of 
these issues, based
on numerical study of simplified ``cosmological-like'' 
simulations, can be found in \cite{sl1, sl2}. We will explore
them further for the specific question of quantification
of errors in this regime in forthcoming publications. 

\noindent
{\bf Acknowledgements}

We thank T. Baertschiger, A. Gabrielli and F. Sylos Labini
for many useful discussions and comments.

\appendix

\section{Eigenmodes in an EdS universe}
\label{appendix-EdSmodefunctions}

The evolution of the displacement field depends, as we have emphasized,
on the cosmological model only through the mode functions $U_n(\bk, t)$
and $V_n(\bk, t)$, which are solutions of Eq.~(\ref{mode-equation})
satisfying the conditions in Eq~.(\ref{normalization}). In the
the case of an EdS universe the former becomes
\be
\label{modes-exp-expl}
\ddot f (\bk,t)+\frac{4}{3t}\dot f (\bk,t)=\frac{2}{3t^2}\epsilon_n(\bk)f_n(\bk,t),
\ee
where, as above, $\epsilon_n(\bk)=-\omega_n^2(\bk)/4\pi G\rho_0$, 
and we have used 
\be
a(t)=\left(\frac{t}{t_0}\right)^{2/3},\qquad 6\pi G\rho_0 t_0^2=1, 
\ee
It follows that 
\bse
\label{uv-exp}
\begin{align}
U_n(\bk,t)=&\tilde\al(\bk)
\left[\al_n^{+}(\bk)\left(\frac{t}{t_0}\right)^{\al_n^{-}(\bk)}+\al_n^{-}(\bk)\left(\frac{t}{t_0}\right)^{-\al_n^{+}(\bk)}\right]\\
V_n(\bk,t)=&\tilde\al(\bk)t_0
\left[\left(\frac{t}{t_0}\right)^{\al_n^{-}(\bk)}-\left(\frac{t}{t_0}\right)^{-\al_n^{+}(\bk)}\right]
\end{align}
\ese
where
\be
\tilde\al(\bk)=\frac{1}{\al_n^{-}(\bk)+\al_n^{+}(\bk)}
\ee
and
\bse
\label{alpha-epsilon}
\begin{align}
&\al_n^{-}(\bk)=\frac{1}{6}\left[\sqrt{1+24\epsilon_n(\bk)}-1\right]\\
&\al_n^{+}(\bk)=\frac{1}{6}\left[\sqrt{1+24\epsilon_n(\bk)}+1\right].
\end{align}
\ese
Note that in the fluid limit we obtain $\alpha_- = 2/3$ and 
$\alpha_+ =1$ for the longitudinal mode ($\epsilon_n = 1$)
and $\alpha_- = 0$ and $\alpha_+ = 1/3$ for the tranverse
modes ($\epsilon_n = 0$).
If $\epsilon_n(\bk)>0$ the solution are a power-law growing mode
and a power-law decaying mode. If $0>\epsilon_n(\bk)>-1/24$, there are two
decaying modes. Finally, if $\epsilon_n(\bk)\leq-1/24$, the solution is
oscillatory and can be written as
\bse
\label{uv-exp-neg}
\begin{align}
U_n(\bk,t)=&\left(\frac{t}{t_0}\right)^{-\frac{1}{6}}\cos\left[\ga_n(\bk)\ln\left(\frac{t}{t_0}\right)\right]\\\nonumber
&+\frac{1}{6\ga_n(\bk)}\left(\frac{t}{t_0}\right)^{-\frac{1}{6}}\sin\left[\ga_n(\bk)\ln\left(\frac{t}{t_0}\right)\right]\\
V_n(\bk,t)=&\frac{t_0}{\ga_n(\bk)}\left(\frac{t}{t_0}\right)^{-\frac{1}{6}}\sin\left[\ga_n(\bk)\ln\left(\frac{t}{t_0}\right)\right]
\end{align}
\ese
where
\be
\ga_n(\bk)=\frac{1}{6}\sqrt{|24\epsilon_n(\bk)+1|},
\ee

\onecolumngrid   


\begin{thebibliography}{37}
\expandafter\ifx\csname natexlab\endcsname\relax\def\natexlab#1{#1}\fi
\expandafter\ifx\csname bibnamefont\endcsname\relax
  \def\bibnamefont#1{#1}\fi
\expandafter\ifx\csname bibfnamefont\endcsname\relax
  \def\bibfnamefont#1{#1}\fi
\expandafter\ifx\csname citenamefont\endcsname\relax
  \def\citenamefont#1{#1}\fi
\expandafter\ifx\csname url\endcsname\relax
  \def\url#1{\texttt{#1}}\fi
\expandafter\ifx\csname urlprefix\endcsname\relax\def\urlprefix{URL }\fi
\providecommand{\bibinfo}[2]{#2}
\providecommand{\eprint}[2][]{\url{#2}}

\bibitem[{\citenamefont{Efstathiou et~al.}(1985)\citenamefont{Efstathiou,
  Davis, Frenk, and White}}]{efstathiou_init}
\bibinfo{author}{\bibfnamefont{G.}~\bibnamefont{Efstathiou}},
  \bibinfo{author}{\bibfnamefont{M.}~\bibnamefont{Davis}},
  \bibinfo{author}{\bibfnamefont{C.~S.} \bibnamefont{Frenk}}, \bibnamefont{and}
  \bibinfo{author}{\bibfnamefont{S.~D.~M.} \bibnamefont{White}},
  \bibinfo{journal}{Astrophys. J. Supp.} \textbf{\bibinfo{volume}{57}},
  \bibinfo{pages}{241} (\bibinfo{year}{1985}).

\bibitem[{\citenamefont{Couchman}(1991)}]{couchman}
\bibinfo{author}{\bibfnamefont{H.~M.~P.} \bibnamefont{Couchman}},
  \bibinfo{journal}{Astrophys. J.} \textbf{\bibinfo{volume}{368}},
  \bibinfo{pages}{L32} (\bibinfo{year}{1991}).

\bibitem[{\citenamefont{Springel et~al.}(2005)}]{springel_05}
\bibinfo{author}{\bibfnamefont{V.}~\bibnamefont{Springel}}
  \bibnamefont{et~al.}, \bibinfo{journal}{Nature}
  \textbf{\bibinfo{volume}{435}}, \bibinfo{pages}{629} (\bibinfo{year}{2005}),
  \eprint{astro-ph/0504097}.

\bibitem[{\citenamefont{Huterer and Takada}(2005)}]{Huterer:2004tr}
\bibinfo{author}{\bibfnamefont{D.}~\bibnamefont{Huterer}} \bibnamefont{and}
  \bibinfo{author}{\bibfnamefont{M.}~\bibnamefont{Takada}},
  \bibinfo{journal}{Astropart. Phys.} \textbf{\bibinfo{volume}{23}},
  \bibinfo{pages}{369} (\bibinfo{year}{2005}), \eprint{astro-ph/0412142}.

\bibitem[{\citenamefont{Joyce and Marcos}(2007)}]{discreteness1_mjbm}
\bibinfo{author}{\bibfnamefont{M.}~\bibnamefont{Joyce}} \bibnamefont{and}
  \bibinfo{author}{\bibfnamefont{B.}~\bibnamefont{Marcos}},
  \bibinfo{journal}{Phys. Rev.} \textbf{\bibinfo{volume}{D75}},
  \bibinfo{pages}{063516} (\bibinfo{year}{2007}), \eprint{astro-ph/0410451}.

\bibitem[{\citenamefont{Braun and Hepp}(1977)}]{braun+hepp}
\bibinfo{author}{\bibfnamefont{W.}~\bibnamefont{Braun}} \bibnamefont{and}
  \bibinfo{author}{\bibfnamefont{K.}~\bibnamefont{Hepp}},
  \bibinfo{journal}{Comm. Math. Phys.} \textbf{\bibinfo{volume}{56}},
  \bibinfo{pages}{101} (\bibinfo{year}{1977}).

\bibitem[{\citenamefont{Spohn}(1991)}]{spohn}
\bibinfo{author}{\bibfnamefont{H.}~\bibnamefont{Spohn}},
  \emph{\bibinfo{title}{Large Scale Dynamics of Interacting Particles}}
  (\bibinfo{publisher}{Springer-Verlag}, \bibinfo{year}{1991}).

\bibitem[{\citenamefont{Scoccimarro}(1998)}]{scoccimarro_transients_98}
\bibinfo{author}{\bibfnamefont{R.}~\bibnamefont{Scoccimarro}},
  \bibinfo{journal}{Mon. Not. R. Astron. Soc.} \textbf{\bibinfo{volume}{299}},
  \bibinfo{pages}{1097} (\bibinfo{year}{1998}).

\bibitem[{\citenamefont{Valageas}(2002)}]{valageas_IC_03}
\bibinfo{author}{\bibfnamefont{P.}~\bibnamefont{Valageas}},
  \bibinfo{journal}{Astron. Astrophys} \textbf{\bibinfo{volume}{385}},
  \bibinfo{pages}{761} (\bibinfo{year}{2002}).

\bibitem[{\citenamefont{Power et~al.}(2003)}]{power_03}
\bibinfo{author}{\bibfnamefont{C.}~\bibnamefont{Power}} \bibnamefont{et~al.},
  \bibinfo{journal}{Mon. Not. Roy. Astron. Soc.}
  \textbf{\bibinfo{volume}{338}}, \bibinfo{pages}{14} (\bibinfo{year}{2003}),
  \eprint{astro-ph/0201544}.

\bibitem[{\citenamefont{Tatekawa and Mizuno}()}]{TatekawaIC_07}
\bibinfo{author}{\bibfnamefont{T.}~\bibnamefont{Tatekawa}} \bibnamefont{and}
  \bibinfo{author}{\bibfnamefont{S.}~\bibnamefont{Mizuno}},
  \emph{\bibinfo{title}{Transients from initial conditions based on lagrangian
  perturbation theory in n-body simulations}},
  \bibinfo{note}{astro-ph/07061334}.

\bibitem[{\citenamefont{Joyce et~al.}(2005)\citenamefont{Joyce, Marcos,
  Gabrielli, Baertschiger, and Sylos~Labini}}]{joyce_05}
\bibinfo{author}{\bibfnamefont{M.}~\bibnamefont{Joyce}},
  \bibinfo{author}{\bibfnamefont{B.}~\bibnamefont{Marcos}},
  \bibinfo{author}{\bibfnamefont{A.}~\bibnamefont{Gabrielli}},
  \bibinfo{author}{\bibfnamefont{T.}~\bibnamefont{Baertschiger}},
  \bibnamefont{and}
  \bibinfo{author}{\bibfnamefont{F.}~\bibnamefont{Sylos~Labini}},
  \bibinfo{journal}{Phys. Rev. Lett.} \textbf{\bibinfo{volume}{95}},
  \bibinfo{pages}{011304} (\bibinfo{year}{2005}), \eprint{astro-ph/0504213}.

\bibitem[{\citenamefont{Marcos et~al.}(2006)\citenamefont{Marcos, Baertschiger,
  Joyce, Gabrielli, and Sylos~Labini}}]{marcos_06}
\bibinfo{author}{\bibfnamefont{B.}~\bibnamefont{Marcos}},
  \bibinfo{author}{\bibfnamefont{T.}~\bibnamefont{Baertschiger}},
  \bibinfo{author}{\bibfnamefont{M.}~\bibnamefont{Joyce}},
  \bibinfo{author}{\bibfnamefont{A.}~\bibnamefont{Gabrielli}},
  \bibnamefont{and}
  \bibinfo{author}{\bibfnamefont{F.}~\bibnamefont{Sylos~Labini}},
  \bibinfo{journal}{Phys. Rev} \textbf{\bibinfo{volume}{D73}},
  \bibinfo{pages}{103507} (\bibinfo{year}{2006}), \eprint{astro-ph/0601479}.

\bibitem[{\citenamefont{Pines}(1963)}]{pines}
\bibinfo{author}{\bibfnamefont{D.}~\bibnamefont{Pines}},
  \emph{\bibinfo{title}{Elementary Excitations in Solids}}
  (\bibinfo{publisher}{Benjamin, New York}, \bibinfo{year}{1963}).

\bibitem[{\citenamefont{Buchert}(1992)}]{buchert2}
\bibinfo{author}{\bibfnamefont{T.}~\bibnamefont{Buchert}},
  \bibinfo{journal}{Mon. Not. R. Astron. Soc.} \textbf{\bibinfo{volume}{254}},
  \bibinfo{pages}{729} (\bibinfo{year}{1992}).

\bibitem[{\citenamefont{Splinter et~al.}(1998)\citenamefont{Splinter, Melott,
  Shandarin, and Suto}}]{splinter}
\bibinfo{author}{\bibfnamefont{R.~J.} \bibnamefont{Splinter}},
  \bibinfo{author}{\bibfnamefont{A.~L.} \bibnamefont{Melott}},
  \bibinfo{author}{\bibfnamefont{S.~F.} \bibnamefont{Shandarin}},
  \bibnamefont{and} \bibinfo{author}{\bibfnamefont{Y.}~\bibnamefont{Suto}},
  \bibinfo{journal}{Astrophys. J.} \textbf{\bibinfo{volume}{497}},
  \bibinfo{pages}{38} (\bibinfo{year}{1998}).

\bibitem[{\citenamefont{Kuhlman et~al.}(1996)\citenamefont{Kuhlman, Melott, and
  Shandarin}}]{kuhlman}
\bibinfo{author}{\bibfnamefont{B.}~\bibnamefont{Kuhlman}},
  \bibinfo{author}{\bibfnamefont{A.~L.} \bibnamefont{Melott}},
  \bibnamefont{and} \bibinfo{author}{\bibfnamefont{S.~F.}
  \bibnamefont{Shandarin}}, \bibinfo{journal}{Astrophys. J.}
  \textbf{\bibinfo{volume}{470}}, \bibinfo{pages}{L41} (\bibinfo{year}{1996}).

\bibitem[{\citenamefont{Melott}(1990)}]{melott_alone}
\bibinfo{author}{\bibfnamefont{A.~L.} \bibnamefont{Melott}},
  \bibinfo{journal}{Comments Astrophys.} \textbf{\bibinfo{volume}{15}},
  \bibinfo{pages}{1} (\bibinfo{year}{1990}).

\bibitem[{\citenamefont{Melott et~al.}(1997)\citenamefont{Melott, Shandarin,
  Splinter, and Suto}}]{melott_all}
\bibinfo{author}{\bibfnamefont{A.~L.} \bibnamefont{Melott}},
  \bibinfo{author}{\bibfnamefont{S.~F.} \bibnamefont{Shandarin}},
  \bibinfo{author}{\bibfnamefont{R.~J.} \bibnamefont{Splinter}},
  \bibnamefont{and} \bibinfo{author}{\bibfnamefont{Y.}~\bibnamefont{Suto}},
  \bibinfo{journal}{Astrophys. J.} \textbf{\bibinfo{volume}{479}},
  \bibinfo{pages}{L79} (\bibinfo{year}{1997}).

\bibitem[{\citenamefont{Diemand
  et~al.}(2004{\natexlab{a}})\citenamefont{Diemand, Moore, and
  Stadel}}]{diemandetal_convergence}
\bibinfo{author}{\bibfnamefont{J.}~\bibnamefont{Diemand}},
  \bibinfo{author}{\bibfnamefont{B.}~\bibnamefont{Moore}}, \bibnamefont{and}
  \bibinfo{author}{\bibfnamefont{J.}~\bibnamefont{Stadel}},
  \bibinfo{journal}{Mon. Not. R. Astron. Soc.} p. \bibinfo{pages}{624}
  (\bibinfo{year}{2004}{\natexlab{a}}).

\bibitem[{\citenamefont{Diemand
  et~al.}(2004{\natexlab{b}})\citenamefont{Diemand, Moore, Stadel, and
  Kazantzidis}}]{diemandetal_2body}
\bibinfo{author}{\bibfnamefont{J.}~\bibnamefont{Diemand}},
  \bibinfo{author}{\bibfnamefont{B.}~\bibnamefont{Moore}},
  \bibinfo{author}{\bibfnamefont{J.}~\bibnamefont{Stadel}}, \bibnamefont{and}
  \bibinfo{author}{\bibfnamefont{S.}~\bibnamefont{Kazantzidis}},
  \bibinfo{journal}{Mon. Not. R. Astron. Soc.} p. \bibinfo{pages}{977}
  (\bibinfo{year}{2004}{\natexlab{b}}).

\bibitem[{\citenamefont{Binney}(2004)}]{binney_discreteness}
\bibinfo{author}{\bibfnamefont{J.}~\bibnamefont{Binney}},
  \bibinfo{journal}{Mon. Not. R. Astron. Soc.} \textbf{\bibinfo{volume}{350}},
  \bibinfo{pages}{939} (\bibinfo{year}{2004}).

\bibitem[{\citenamefont{Hamana et~al.}(2002)\citenamefont{Hamana, Yoshida, and
  Suto}}]{discreteness-hamana}
\bibinfo{author}{\bibfnamefont{T.}~\bibnamefont{Hamana}},
  \bibinfo{author}{\bibfnamefont{N.}~\bibnamefont{Yoshida}}, \bibnamefont{and}
  \bibinfo{author}{\bibfnamefont{Y.}~\bibnamefont{Suto}},
  \bibinfo{journal}{Astrophys. J.} \textbf{\bibinfo{volume}{568}},
  \bibinfo{pages}{455} (\bibinfo{year}{2002}).

\bibitem[{\citenamefont{Baertschiger et~al.}(2002)\citenamefont{Baertschiger,
  Joyce, and Sylos~Labini}}]{Baertschiger:2002tk}
\bibinfo{author}{\bibfnamefont{T.}~\bibnamefont{Baertschiger}},
  \bibinfo{author}{\bibfnamefont{M.}~\bibnamefont{Joyce}}, \bibnamefont{and}
  \bibinfo{author}{\bibfnamefont{F.}~\bibnamefont{Sylos~Labini}},
  \bibinfo{journal}{Astrophys. J.} \textbf{\bibinfo{volume}{581}},
  \bibinfo{pages}{L63} (\bibinfo{year}{2002}), \eprint{astro-ph/0203087}.

\bibitem[{\citenamefont{Gotz and Sommer-Larsen}(2003)}]{gotz+sommerlarsen_WDM}
\bibinfo{author}{\bibfnamefont{M.}~\bibnamefont{Gotz}} \bibnamefont{and}
  \bibinfo{author}{\bibfnamefont{J.}~\bibnamefont{Sommer-Larsen}},
  \bibinfo{journal}{Astrophys.Space Sci. 284} \textbf{\bibinfo{volume}{284}},
  \bibinfo{pages}{341} (\bibinfo{year}{2003}).

\bibitem[{\citenamefont{Wang and White}()}]{wang+white_HDM}
\bibinfo{author}{\bibfnamefont{J.}~\bibnamefont{Wang}} \bibnamefont{and}
  \bibinfo{author}{\bibfnamefont{S.}~\bibnamefont{White}},
  \emph{\bibinfo{title}{Discreteness effects in simulations of hot/warm dark
  matter}}, \eprint{astro-ph/0702575}.

\bibitem[{\citenamefont{Marcos}()}]{marcos07}
\bibinfo{author}{\bibfnamefont{B.}~\bibnamefont{Marcos}},
  \emph{\bibinfo{title}{Particle linear theory on a self-gravitating perturbed
  cubic bravais lattice}}, \bibinfo{note}{in preparation}.

\bibitem[{\citenamefont{Zeldovich}(1970)}]{zeldovich_70}
\bibinfo{author}{\bibfnamefont{Y.~B.} \bibnamefont{Zeldovich}},
  \bibinfo{journal}{Astron. Astrophys.} \textbf{\bibinfo{volume}{5}},
  \bibinfo{pages}{84} (\bibinfo{year}{1970}).

\bibitem[{\citenamefont{Bertschinger}(1995)}]{bertschingercode}
\bibinfo{author}{\bibfnamefont{E.}~\bibnamefont{Bertschinger}},
  \emph{\bibinfo{title}{Cosmics: Cosmological initial conditions and microwave
  anisotropy codes}} (\bibinfo{year}{1995}),
  \urlprefix\url{http://www.citebase.org/cgi-bin/citations?id=oai:arXiv.org:as%
tro-ph/9506070}.

\bibitem[{\citenamefont{Schneider and Bartelmann}(1995)}]{sb95}
\bibinfo{author}{\bibfnamefont{P.}~\bibnamefont{Schneider}} \bibnamefont{and}
  \bibinfo{author}{\bibfnamefont{M.}~\bibnamefont{Bartelmann}},
  \bibinfo{journal}{Mon. Not. R. Astron. Soc.} \textbf{\bibinfo{volume}{273}},
  \bibinfo{pages}{475} (\bibinfo{year}{1995}).

\bibitem[{\citenamefont{Gabrielli}(2004)}]{andrea}
\bibinfo{author}{\bibfnamefont{A.}~\bibnamefont{Gabrielli}},
  \bibinfo{journal}{Phys. Rev.} \textbf{\bibinfo{volume}{E70}},
  \bibinfo{pages}{066131} (\bibinfo{year}{2004}), \eprint{cond-mat/0409594}.

\bibitem[{\citenamefont{Binney and Tremaine}(1994)}]{binney}
\bibinfo{author}{\bibfnamefont{J.}~\bibnamefont{Binney}} \bibnamefont{and}
  \bibinfo{author}{\bibfnamefont{S.}~\bibnamefont{Tremaine}},
  \emph{\bibinfo{title}{Galactic Dynamics}} (\bibinfo{publisher}{Princeton
  University Press}, \bibinfo{year}{1994}).

\bibitem[{\citenamefont{Yamaguchi et~al.}(2004)\citenamefont{Yamaguchi,
  Barr\'e, Bouchet, Dauxois, and Ruffo}}]{yamaguchi_etal_04}
\bibinfo{author}{\bibfnamefont{Y.~Y.} \bibnamefont{Yamaguchi}},
  \bibinfo{author}{\bibfnamefont{J.}~\bibnamefont{Barr\'e}},
  \bibinfo{author}{\bibfnamefont{F.}~\bibnamefont{Bouchet}},
  \bibinfo{author}{\bibfnamefont{T.}~\bibnamefont{Dauxois}}, \bibnamefont{and}
  \bibinfo{author}{\bibfnamefont{S.}~\bibnamefont{Ruffo}},
  \bibinfo{journal}{Physica A} \textbf{\bibinfo{volume}{337}},
  \bibinfo{pages}{36} (\bibinfo{year}{2004}), \eprint{cond-mat/0312480}.

\bibitem[{\citenamefont{White}(1993)}]{white_leshouches}
\bibinfo{author}{\bibfnamefont{S.}~\bibnamefont{White}},
  \emph{\bibinfo{title}{Lectures given at les houches}} (\bibinfo{year}{1993}),
  \eprint{astro-ph/9410043}.

\bibitem[{\citenamefont{Joyce et~al.}()\citenamefont{Joyce, Marcos, and
  Baertschiger}}]{preIC_07}
\bibinfo{author}{\bibfnamefont{M.}~\bibnamefont{Joyce}},
  \bibinfo{author}{\bibfnamefont{B.}~\bibnamefont{Marcos}}, \bibnamefont{and}
  \bibinfo{author}{\bibfnamefont{T.}~\bibnamefont{Baertschiger}},
  \emph{\bibinfo{title}{Towards a quantification of discreteness error in the
  non-linear regime of cosmological $n$-body simulations}}, \bibinfo{note}{in
  preparation}.

\bibitem[{\citenamefont{Baertschiger
  et~al.}(2007{\natexlab{a}})\citenamefont{Baertschiger, Joyce, Gabrielli, and
  Sylos~Labini}}]{sl1}
\bibinfo{author}{\bibfnamefont{T.}~\bibnamefont{Baertschiger}},
  \bibinfo{author}{\bibfnamefont{M.}~\bibnamefont{Joyce}},
  \bibinfo{author}{\bibfnamefont{A.}~\bibnamefont{Gabrielli}},
  \bibnamefont{and}
  \bibinfo{author}{\bibfnamefont{F.}~\bibnamefont{Sylos~Labini}},
  \bibinfo{journal}{Phys. Rev.} \textbf{\bibinfo{volume}{E75}},
  \bibinfo{pages}{021113} (\bibinfo{year}{2007}{\natexlab{a}}),
  \eprint{cond-mat/0607396}.

\bibitem[{\citenamefont{Baertschiger
  et~al.}(2007{\natexlab{b}})\citenamefont{Baertschiger, Joyce, Gabrielli, and
  Sylos~Labini}}]{sl2}
\bibinfo{author}{\bibfnamefont{T.}~\bibnamefont{Baertschiger}},
  \bibinfo{author}{\bibfnamefont{M.}~\bibnamefont{Joyce}},
  \bibinfo{author}{\bibfnamefont{A.}~\bibnamefont{Gabrielli}},
  \bibnamefont{and}
  \bibinfo{author}{\bibfnamefont{F.}~\bibnamefont{Sylos~Labini}},
  \bibinfo{journal}{Phys. Rev.} \textbf{\bibinfo{volume}{E76}},
  \bibinfo{pages}{011116} (\bibinfo{year}{2007}{\natexlab{b}}),
  \eprint{cond-mat/0612594}.

\end{thebibliography}

\end{document}